\documentclass[a4paper,fleqn,usenatbib]{mnras}

\usepackage{graphicx}	
\usepackage{amssymb}	
\usepackage{natbib}
\usepackage{soul}
\bibpunct{(}{)}{;}{a}{}{,}
\usepackage{graphicx}
\usepackage{multirow}

\usepackage{graphicx}
\usepackage{txfonts}
\usepackage{natbib}
\usepackage{xspace}
\usepackage{dcolumn}
\usepackage{longtable}
\usepackage{lscape}
\usepackage{soul}        

\usepackage{rotating}
\usepackage{afterpage}

\newcommand{\MJup}{M$_{\mathrm{Jup}}$\xspace}

\newcommand{\MSun}{M$_{\odot}$\xspace}

\newcommand{\as}{\hbox{$^{\prime\prime}$}\xspace}
\def\tablebib#1{\par\vspace*{2ex}%
 \parbox{\hsize}{\leftskip0pt\rightskip0pt
 {\noindent\small\textbf{References.}~#1\par}}}

\title[Imaging RV planets with SPHERE]{Imaging radial velocity planets with SPHERE}
\author[A. Zurlo et al.]{A. Zurlo$^{1,2,3}$\thanks{E-mail:
    alice.zurlo@mail.udp.cl}, D. Mesa$^{4,5}$, S. Desidera$^{5}$, S. Messina$^{6}$, R. Gratton$^{5}$,  C. Moutou$^{3}$, J.-L. Beuzit$^{7}$, \newauthor
  B. Biller$^{8}$, A. Boccaletti$^{9}$, M. Bonavita$^{8}$, M. Bonnefoy$^{7}$, T. Bhowmik$^{9}$, W. Brandner$^{10}$, \newauthor E. Buenzli$^{11}$, G. Chauvin$^{7}$, M. Cudel$^{9}$, V. D'Orazi$^{5}$, M. Feldt$^{10}$, J. Hagelberg$^{12}$, M. Janson$^{10,13}$, \newauthor A-M. Lagrange$^{7}$, M. Langlois$^{3,14}$, J. Lannier$^{7}$, B. Lavie$^{12}$, C. Lazzoni$^{5}$, 
A.-L. Maire$^{10}$, \newauthor M. Meyer$^{15}$, D. Mouillet$^{7}$, S. Peretti$^{12}$, C. Perrot$^{9}$, 
P.J. Potiron$^{16}$, G. Salter$^{3}$, T. Schmidt$^{9}$, \newauthor E. Sissa$^{5}$, A. Vigan$^{3}$, A. Delboulb\'e$^{7}$, C. Petit$^{17}$, J. Ramos$^{10}$, F. Rigal$^{18}$, S. Rochat$^{7}$  \\  \\
$^{1}$N\'ucleo de Astronom\'ia, Facultad de Ingenier\'ia y Ciencias, Universidad Diego Portales, Av. Ejercito 441, Santiago, Chile\\
      $^{2}$Escuela de Ingenier\'ia Industrial, Facultad de Ingenier\'ia y Ciencias, Universidad Diego Portales, Av. Ejercito 441, Santiago, Chile \\
      $^{3}$Aix Marseille Universit\'e, CNRS, LAM - Laboratoire d'Astrophysique de Marseille, UMR 7326, 13388, Marseille, France  \\
 $^{4}$INCT, Universidad De Atacama, calle Copayapu 485, Copiap\'{o}, Atacama, Chile \\
$^{5}$INAF-Osservatorio Astronomico di Padova, Vicolo dell'Osservatorio 5, Padova, Italy, 35122-I \\
   $^{6}$INAF-Osservatorio Astrofisico di Catania, Via S. Sofia 78, 95123, Catania, Italy \\
  $^{7}$Universit\'e Grenoble Alpes, IPAG, F-38000 Grenoble, France \\
$^{8}$Institute for Astronomy, The University of Edinburgh, Royal Observatory, Blackford Hill Edinburgh EH9 3HJ U.K. \\
$^{9}$LESIA, Observatoire de Paris, Universit\'e PSL, CNRS, Sorbonne Universit\'e, Univ. Paris Diderot, Sorbonne Paris Cit\'e, 5 place Jules Janssen, 92195 Meudon, France  \\
$^{10}$Max-Planck-Institut f\"ur Astronomie, K\"onigstuhl 17, 69117 Heidelberg, Germany\\
$^{11}$ETH Zurich, Institute for Particle Physics and Astrophysics, Wolfgang-Pauli-Strasse 27, 8093 Zurich, Switzerland\\
$^{12}$Observatoire de Gen\`eve, University of Geneva, 51 Chemin des Maillettes, 1290, Versoix, Switzerland\\
$^{13}$Department of Astronomy, Stockholm University, AlbaNova University Center, 106 91 Stockholm, Sweden \\
$^{14}$CRAL, UMR 5574, CNRS, Universit\'e Lyon 1, 9 avenue Charles Andr\'e, 69561 Saint Genis Laval Cedex, France \\
$^{15}$Department of Astronomy, University of Michigan, 1085 S. University Ave, Ann Arbor, MI 48109-1107, USA \\
$^{16}$Universit\'e Nice-Sophia Antipolis, CNRS, Observatoire de la C\^ote d'Azur, Laboratoire J.-L. Lagrange, CS 34229, 06304 Nice cedex 4, France\\ $^{17}$DOTA, ONERA, Universit\'e Paris Saclay, F-91123, Palaiseau France \\
$^{18}$Anton Pannekoek Institute for Astronomy, Science Park 904, NL-1098 XH Amsterdam, The Netherlands\\}
\begin{document}

\date{Accepted 2018 July 3. Received 2018 July 2; in original form 2018 May 14}

\pagerange{\pageref{firstpage}--\pageref{lastpage}} \pubyear{}

\maketitle

\label{firstpage}
\begin{abstract}
  We present observations with the planet finder SPHERE of a selected sample of the most promising radial velocity (RV) companions for high-contrast imaging.  Using a Monte Carlo simulation to explore all the possible inclinations of the orbit of wide RV companions, we identified the systems with companions that could potentially be detected with SPHERE. We found the most favorable RV systems to observe are : HD\,142, GJ\,676, HD\,39091, HIP\,70849, and HD\,30177 and carried out observations of these systems during SPHERE Guaranteed Time Observing (GTO).
 To reduce the intensity of the starlight and reveal faint companions,  we used Principle Component Analysis (PCA) algorithms alongside angular and spectral differential imaging. 
We injected synthetic planets with known flux to evaluate the self-subtraction caused by our data reduction and to determine the 5$\sigma$ contrast in the J band $vs$ separation for our reduced images.  We estimated the upper limit on detectable companion mass around the selected stars from the contrast plot obtained from our data reduction.
     Although our observations enabled contrasts larger than 15 mag at a few tenths of arcsec from the host stars, we detected no planets. However, we were able to set upper mass limits around the stars using AMES-COND evolutionary models. We can exclude the presence of companions more massive than 25-28 \MJup around these stars, confirming the substellar nature of these RV companions.  

\end{abstract}


\section{Introduction}
So far, 751 planets have been discovered with the radial velocity (RV) technique\footnote{\url{www.exoplanet.eu}}. 
As stellar activity also produces radial velocity variability and can mimic planet signals, these objects' hosts are old and non-active stars.
This method can detect both close planets and massive long-period objects, but, due to the unknown inclination of the detected companions, 
 only the minimum mass can be determined. The measured parameter is the mass of the companion multiplied by the sine of the inclination of its orbit, $M\sin i$. As the inclination is unknown from the RV measurements alone, the real mass of the object cannot be directly measured using this technique alone.
Combining RV and high-contrast imaging measurements of the same companion allows us to constrain 
the companion orbit and thus measure its dynamical mass \citep[see e.g.,][]{boden2006}, providing a crucial benchmark for 
evolutionary models of substellar objects \citep[e.g.,][]{baraffe2015}.

In the last years, a number of 
surveys have been conducted to image previously detected RV companion objects on wide orbits,  
for instance, direct imaging observations of targets with RV
drifts from HARPS and CORALIE employing VLT/NACO \citep{hagelberg2010}, the TREND imaging survey of targets
with known trends in the RV \citep[see, e.g.,][]{crepp2012} and the NICI follow-up survey on long period RV 
targets \citep{salter2014}. So far no planetary companion has both a measurement of dynamical mass and an estimated mass from evolutionary models.
The failure of previous attempts is due to the small apparent separation between the planet and
the star. {The dynamical mass of the brown dwarfs HR\,7672\,B \citep{2002ApJ...571..519L} and HD\,4747\,B \citep{2016ApJ...831..136C} has been successfully measured via the coupling of the direct imaging and the RV technique \citep{2012ApJ...751...97C, 2018ApJ...853..192C, 2018arXiv180505645P}.} The combinations of RV and direct imaging have allowed a better constrain on the atmospheric properties of the companion as the constrain on the mass helps to solve degeneracies in the atmospheric models. Since planets with longer and longer periods (and separations) are being discovered by RV surveys
extending over longer time spans, repeating this attempt with the latest discoveries offers a higher probability of success 
relative to previous attempts.  Constraining mass, age, and chemical composition in this manner presents a unique opportunity to calibrate theoretical evolutionary models that are fundamental in describing disk instability (DI) companions \citep{2002ApJ...567L.133P}.

With this aim, we selected a sample of targets with known companion objects found via the RV method that could potentially be
imaged with the Spectro-Polarimetric High-contrast Exoplanet REsearch (SPHERE) instrument \citep{beuzit2006} at the Very Large Telescope (VLT) on Cerro Paranal. 
SPHERE is a highly specialized instrument dedicated to high contrast imaging, built by a wide consortium of European laboratories. It is based on the SAXO extreme adaptive optics system \citep{Fusco:06, 2014SPIE.9148E..0OP, doi:10.1117/12.2056352}, with a 41$\times$41 actuators wavefront control, pupil stabilization, differential tip-tilt control and employs stress polished toric mirrors for beam transportation \citep{2012A&A...538A.139H}. Several coronagraphic devices for stellar diffraction suppression are provided, including apodized Lyot coronagraphs \citep{2005ApJ...618L.161S,2011ExA....30...59G} and achromatic four-quadrants phase masks \citep{Boc08}. The instrument is equipped with three science channels: the differential imaging camera \citep[IRDIS,][]{Do08}, an integral field spectrograph \citep[IFS,][]{Cl08} and the Zimpol rapid-switching imaging polarimeter \citep[ZIMPOL,][]{Th08}. The system includes a dedicated data pipeline capable of applying basic reduction steps as well as higher-level differential imaging analysis procedures \citep{2008SPIE.7019E..39P}.

Since old planetary mass objects at large separation from the host star are cool and
extremely faint at optical wavelengths, in this paper we utilized IRDIS and IFS, which operate in the IR.
These instruments have been designed to be particularly sensitive to the detection of T dwarf companions, the spectral type that we expect for radial velocity companions.  
The performance of these two instruments has been widely presented in recent results as e.g. \citet{2016A&A...587A..55V,2016A&A...587A..56M, 2016A&A...587A..57Z, 2016A&A...587A..58B}. Based on these results, we expected to reach a contrast of 10$^{-6}$ at a separation of 0\farcs5 with IFS, which would allow us to detect objects of several Jupiter masses around old > Gyr stars. \par
We performed a Monte Carlo simulation, presented in Sec.~\ref{s:sel}, to select the most promising RV companions observable with SPHERE. We found five companions with a non-null detection probability: HD\,142\,Ac, GJ\,676\,b and c, HIP\,70849\,b, and HD\,30177\,b. We also added the star HD\,39091, which has constraints from astrometry rather than RV but is still an intriguing target. These targets have been observed during the GTO campaign in 2014-2017.



We present the selection criteria for our RV sample observed with SPHERE in Section~\ref{s:sel}.
Observations and data analysis are described in Section~\ref{s:obs}; followed by the analysis of each individual target: HD\,142 (Sec.~\ref{s:HD142}), GJ\,676 (Sec.~\ref{s:GJ}), HD\,39091 (Sec.~\ref{s:HD39}), HIP\,70849 (Sec.~\ref{s:HIP}), and HD\,30177 (Sec.~\ref{s:HD30}) and Conclusions (Sec.~\ref{conclusion}).

\section[]{Selection of the RV targets}
\label{s:sel}
In 2014, we selected a sample of long-period RV planets to include in SPHERE GTO.
The input sample was composed of all RV planets with a projected semi-major axis (as listed in the
input catalogues) from the host star larger than 0\farcs1 and with a declination lower than 40
degrees (to be observable from the VLT). The selected minimum separation was chosen to match 
the projected radius of the smallest apodized Lyot coronagraph available with SPHERE, which is 145 mas. Only planets with known
parameters of the orbit have been included, while objects with radial velocity drifts have been excluded.
Using a Monte Carlo simulation we explored the intrinsic magnitude of each companion
and its projected separation from its host star as a function of its unknown orbital inclination. The Monte Carlo simulation explores all the possible inclinations and values of the longitude of ascending node, $\Omega_{node}$,  while the other orbital parameters are fixed by the orbital solution.   
The projected separation between each star and its companion at a given epoch is calculated with the Thiele-Innes
formalism \citep[e.g.,][]{1960pdss.book.....B}. Two epochs have been explored: the apoastron passage, which in the majority of the cases is the moment where the planet is at its farthest position, and the year 2014, when the GTO started.

Several objects within the sample have constraints on the inclination from astrometry
\citep[see, e.g.,][]{2011A&A...527A.140R} that were included in the simulation. The simulation explores different inclinations of the orbit, providing the value of the true mass
of the planet for each inclination.  We used the AMES-COND models \citep{2003IAUS..211..325A} to estimate
the intrinsic magnitude of each planet given its true mass.

\newpage

Because all the objects under 
consideration are older than $\sim$ 1 Gyr, we do not
expect the choice of the model to be crucial, as at ages $\geq$1 Gyr, models with different values of the initial entropy converge \citep[see, e.g.][]{2007ApJ...655..541M}.  The magnitude strongly depends on the age of the system, thus the error bars are correlated with the uncertainty on the age of the host star.  For each target, we calculated the contrast limit reachable with IFS using the official ESO exposure time calculator (ETC), assuming one hour of exposure time, as requested for the observations. We considered only the IFS curve as it reaches deeper contrasts at small separations compared to IRDIS.

\begin{center}
	\begin{figure}
	\centering
	\includegraphics[width=0.5\textwidth]{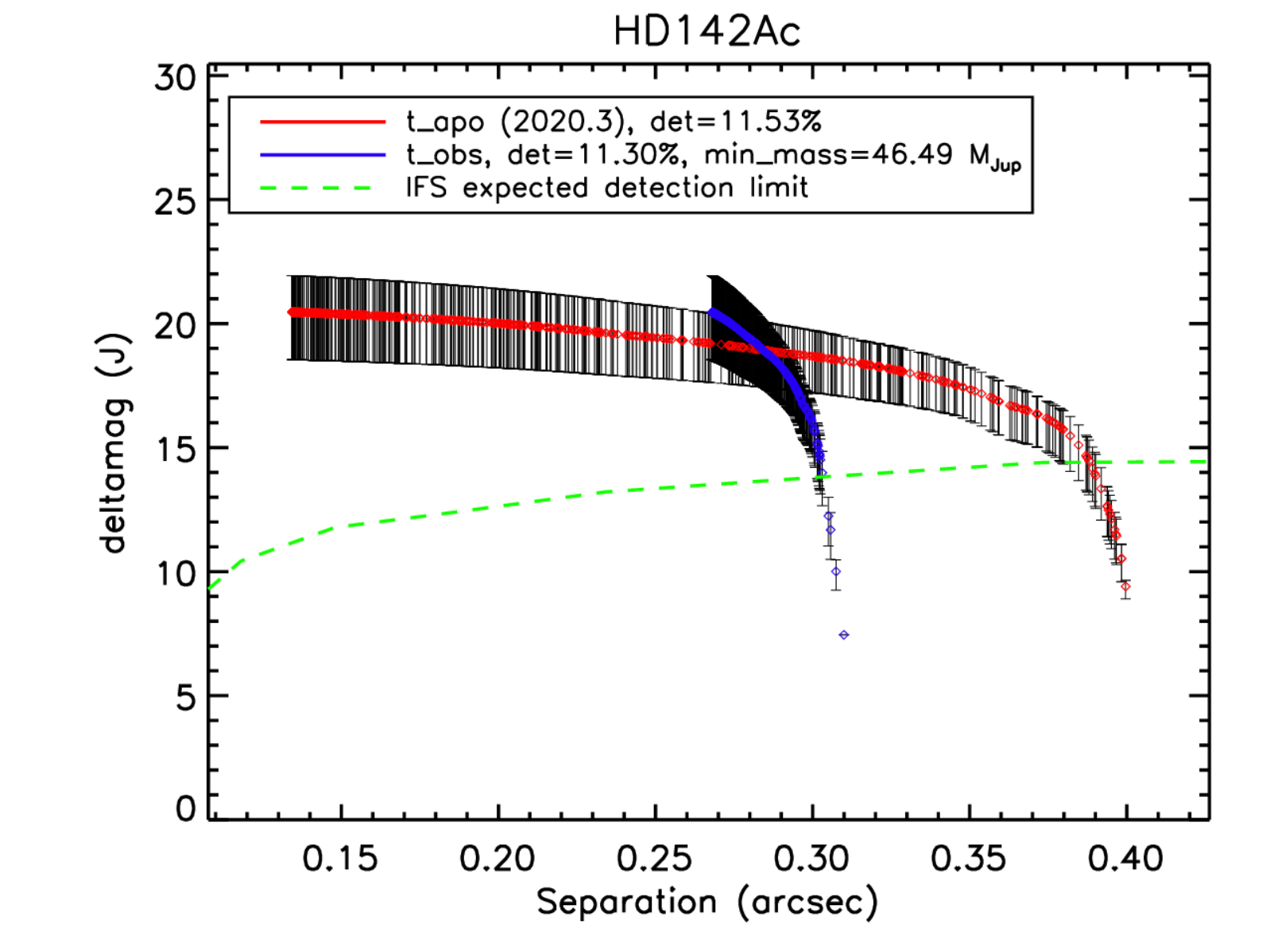}
	\caption{Contrast in $J$ band of IFS for the planet HD\,142\,Ac. The blue dots represent the expected contrast of the planet for the observation date (October 2014) calculated with the Monte Carlo simulation with the inclination as a free parameter. The red dots represent the expected contrast of the planet during the apoastron passage. The error bars are related to the uncertainty on the age of the system. The green dashed line shows the predicted detection limit calculated with the ETC for IFS. In the legend of this Figure we provide the detection probabilities at both epochs considered. As we will show in the next Sections, the expected contrast limits are pessimistic.}
	\label{f:hd142}
        \end{figure}
	\end{center}

In Figure~\ref{f:hd142}, we show an example of the output of the Monte Carlo simulation. The green dashed curve represents the expected contrast achievable with IFS in $J$ band, for a 1h observation as calculated using the ETC.
The red dots in Figure~\ref{f:hd142}
represent the calculated contrast in $J$ magnitude for different inclinations of the planetary orbit resulting in different planetary masses as well as different separations at the
foreseen passage of the planet at the apoastron (in this case at the beginning of 2020). The blue dots represent the same calculation made for the period
of the scheduled SPHERE observations. We note that the ETC tends to be conservative -- the actual contrast limits we can reach are $\sim$ 2 magnitudes better. The minimum mass listed in the plot is the minimum possible mass of the companion which would be detectable with SPHERE.


As we do not know the true inclination of each system, we calculate a detection probability for each planet by inferring the probability that the mass is larger than the minimum mass for detection. To this aim, we used the {\it a posteriori} probability distribution of $\sin{i}$ suggested by \citet{2011ApJ...739...26H}. The details of the detection probability calculation 
are found in Appendix~\ref{a:prob}. We selected for observation all the objects that have a detection probability higher than 5\%. The most promising targets observable with
SPHERE had detection probabilities spanning from $\sim$6\% to $\sim$20\%: HD\,142, GJ\,676, HD\,39091, HIP\,70849, and HD\,30177. We have continued to update all the radial velocity information regarding the orbits of these objects after the first selection made in 2014.

\section{Observations of the targets}
\label{s:obs}

As part of the SPHERE GTO, we observed five promising targets with detection probabilities $>$ 5$\%$: HD\,142, GJ\,676, HD\,39091, HIP\,70849, and HD\,30177. We adopted the same observing strategy for all objects: 
IRDIFS observations, with IRDIS in $H2H3$ bands \citep[$H2$ centered at 1.587 $\mu$m and $H3$ centered at 1.667 $\mu$m][]{vigan2010} and IFS in $YJ$ mode \citep[0.95-1.35 $\mu$m][]{zurlo2014}, in pupil stabilized mode during meridian passage, to take advantage of Angular Differential Imaging \citep[ADI;][]{2006ApJ...641..556M}. The stars HD\,142, HIP\,70849, and HD\,39091, and HD\,30177 have been observed once, while GJ\,676 was observed twice in order to obtain astrometric followup for several candidate companions identified during the initial observing epoch (all candidate companions were background objects, see Appendix~\ref{a:cand}). More details on each observation can be found in Table~\ref{t:obs}.

\begin{table}
 \centering
\caption[]{Observation details for the RV special GTO targets.}
\label{t:obs}
\begin{tabular}{lccc}
\hline
\hline
Target       & Epoch & Seeing (\as)   &  $\Delta$PA (deg)  \\
\hline
HD\,142 & 2014-10-13 & 0.9  &35 \\
GJ\,676 & 2015-05-06 & 0.7 & 34 \\
&   2016-05-30  & 1.0 & 38 \\
HD\,39091 & 2015-12-20  & 0.6 & 24  \\
HIP\,70849 & 2015-05-05 & 0.9 & 34 \\
HD\,30177 & 2017-11-03 & 0.5 & 35  \\

\hline
\end{tabular}
\end{table}


For all the datasets, data from each of the IRDIS dual-band filters images, $H2$ and $H3$, were reduced separately. After background subtraction and flat-fielding, we suppressed the residual speckle noise using ADI with the KLIP algorithm \citep{2012ApJ...755L..28S}. The procedure is the same used to reduce the data for HR\,8799, and we refer to the related paper \citet{2016A&A...587A..57Z} for the full description.

\begin{center}
	\begin{figure}
	\centering
	\includegraphics[width=0.4\textwidth]{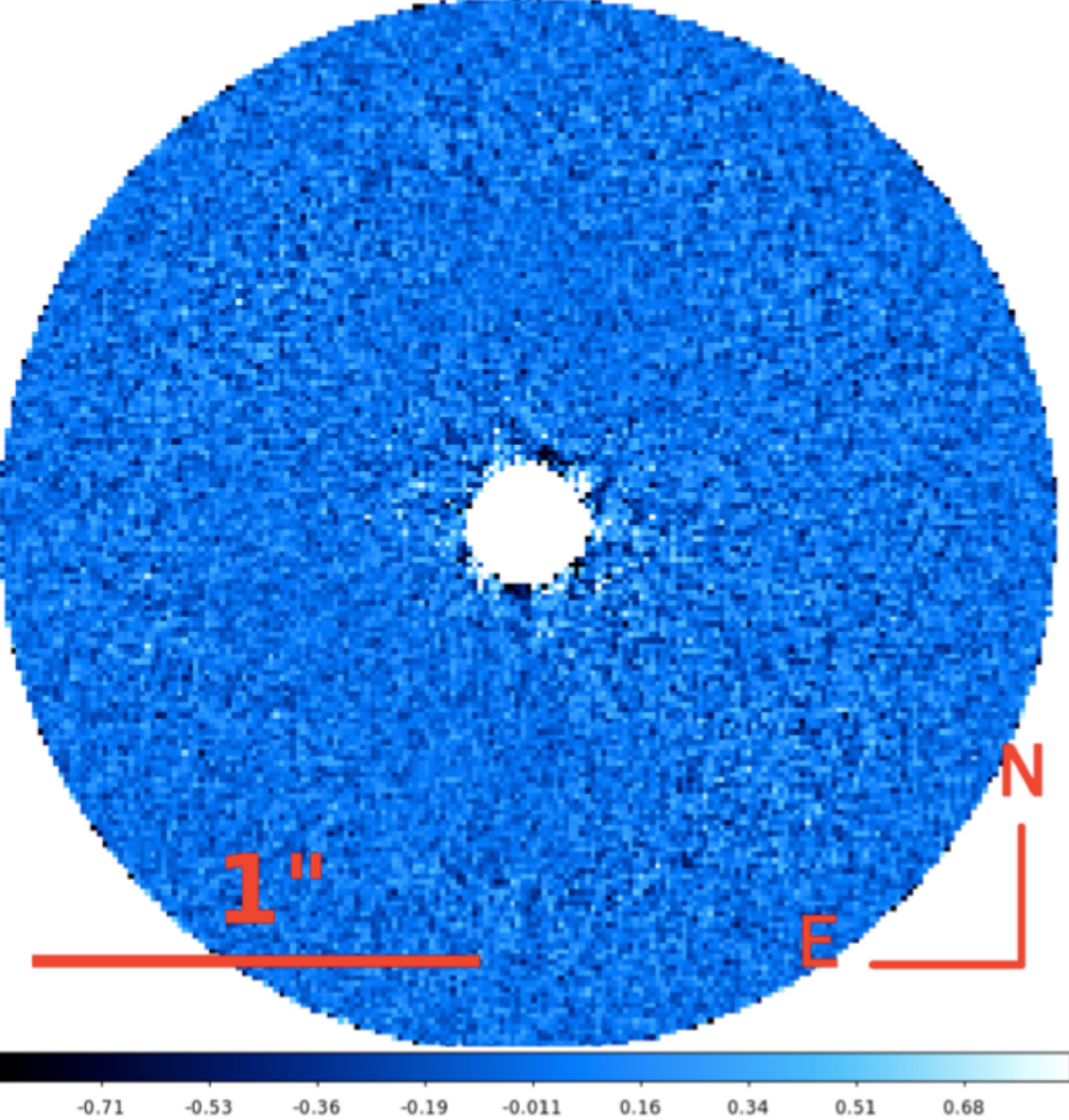}
	\caption{ADI reduction of SPHERE/IRDIS $H2H3$ channels for the object HD\,142\,A (the mean images of the two filters is shown). In the image, no candidate candidates with S/N above 5$\sigma$ are detected. The stellar companion is not visible as only the inner part of the FOV is shown. The scale indicates the contrast.}
 	
	\label{f:fi_hd142}
        \end{figure}
	\end{center}

As an example of the typical results from this procedure, the reduced IRDIS image of HD\,142 is shown in Figure~\ref{f:fi_hd142}. We also performed ADI + Spectral Differential Imaging \citep[SDI,][]{2007ApJ...670.1367L}, obtaining similar results. The contrast reached is not as deep as that reached by IFS at close-in separations, thus we adopt the IFS results for the following analysis. \par
\begin{center}
	\begin{figure*}
	\centering
         \includegraphics[width=0.8\textwidth]{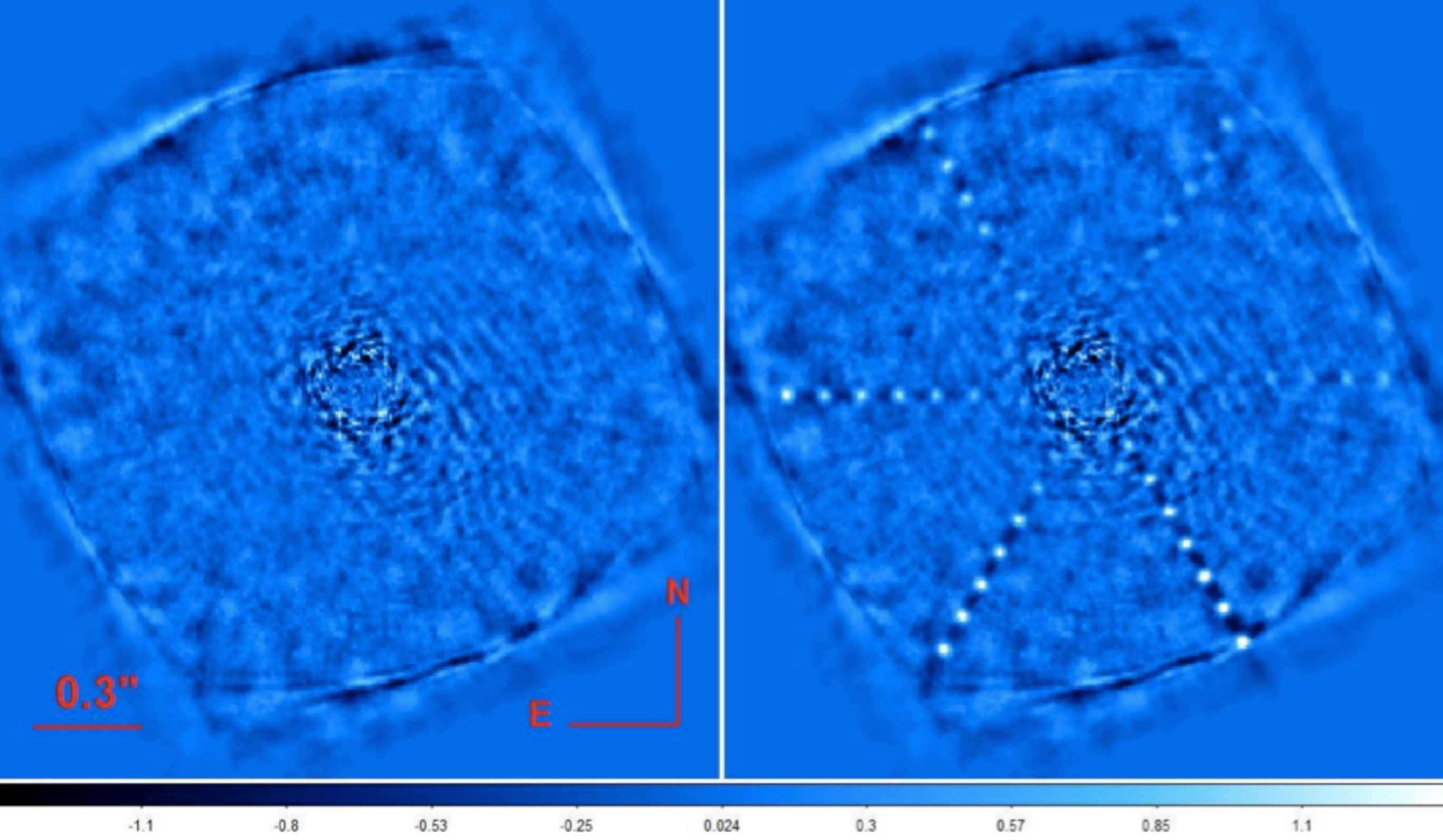}
	\caption{{\it Left:} Result of the PCA SVD-based procedure for the IFS data for HD\,142\,A, after median combination along IFS' 39 spectral channels and using 100 principle components in the
data reduction. {\it Right:} Final result of the PCA procedure for the IFS data for HD\,142\,A with simulated planets inserted at separations ranging between
        0\farcs2 and 0\farcs8 and with log(contrast) ranging from -6.1 to -5.6 with a step of 0.1 (increasing counter-clockwise). 
        Again, 100 principal components were used in the data reduction.}
 		\label{f:ifs_final}
        \end{figure*}
	\end{center}

The IFS data were reduced with the data reduction
and handling (DRH) recipes \citep{2008SPIE.7019E..39P}. The DRH produces a calibrated datacube for each input raw frame. The speckle subtraction on these datacubes was performed with KLIP and with a second principal component
analysis (PCA) method utilizing a single value decomposition (SVD) algorithm adapted to the case of the SPHERE IFS \citep[see][]{2015A&A...576A.121M}. The median collapsed final result is displayed in Figure~\ref{f:ifs_final}, left side, using 100 principal components.  To select the appropriate number of principal components to apply, we performed PCA using various numbers of components and evaluated the final contrast after taking into account self-subtraction. In this manner we verified that 100 principal components maximizes the achieved contrast in these data
relative to more or fewer principal components.

\newpage

\begin{center}
	\begin{figure}
	\centering
	\includegraphics[width=0.5\textwidth]{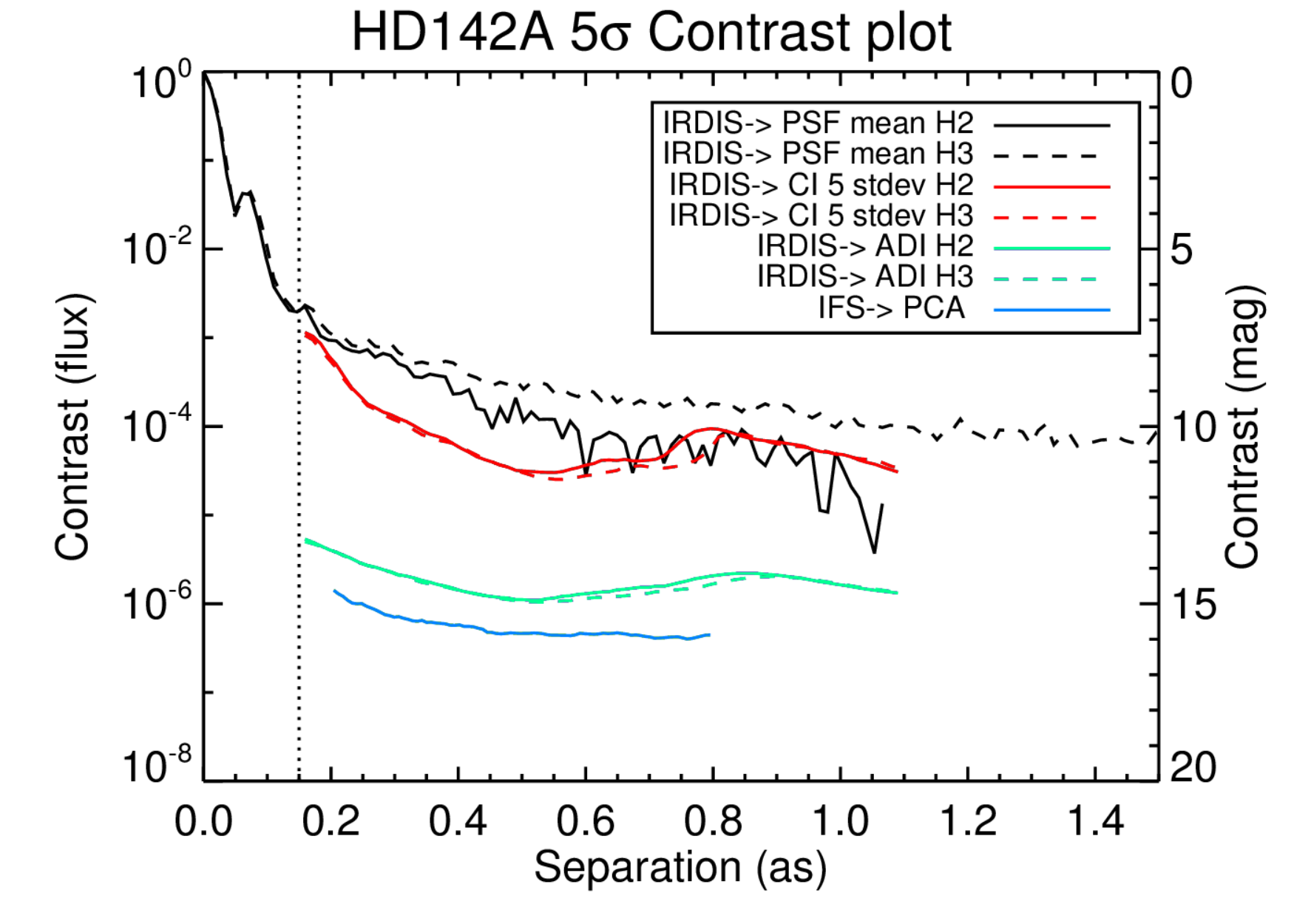}
	\caption{5$\sigma$ residual noise levels as a function of separation for IRDIFS observations of HD\,142\,A.  The mean azimuthal profile of the off-axis PSF (black), the coronagraphic profile (red) are shown for the two IRDIS channels $H2$ (continuous line) and $H3$ (dashed line). The IRDIS contrast plot after the implementation of ADI using the KLIP algorithm is shown in green both for the H2 filter (solid line) and the H3 filter (dashed line). The light blue lines give the contrast obtained for IFS using the PCA algorithm. The solid line gives the contrast obtained injecting a T5-type spectrum planet while the dashed line gives the contrast obtained injecting a L0-type spectrum planet.}
	\label{f:con_hd142}
        \end{figure}
\end{center}

The contrast limits for both instruments have been calculated using the same method as in \citet{2015A&A...576A.121M}
and are shown in Figure~\ref{f:con_hd142}.
Self-subtraction is the most important issue to consider when calculating the final contrast
obtained with high-contrast imaging methods. To properly account for it, we performed a number of different tests with synthetic
planets.  First, we inserted synthetic planets (created by scaling the off-axis PSF) at a variety of separations 
from the central star with star-planet contrasts very close to the detection limit estimated without 
accounting for self-subtraction. In this way we were able to 
evaluate the self-subtraction caused by our method and subsequently to correct the contrast previously calculated.
Second, we introduced simulated planets with contrast just above and below the contrast 
limit calculated with the previous step.   Checking that the simulated planets inserted above the detection limit were indeed 
detected allows us to confirm the contrast curve obtained with the method described above.  An example of the results
of this test is provided in Figure~\ref{f:ifs_final}, right panel.
To define contrast curves for companions with both L and T spectral types, we injected simulated planets both 
with L0-type spectrum \citep{testi2001} and 
T5-type spectrum \citep{burgasser2004}.  We note that we perform these detailed planet simulation and retrieval tests only 
for the IFS data, as the IFS data provide deeper contrasts at the separations of interest compared to the IRDIS data. The transmission of the coronagraph at close separations has been taken into account. 

\begin{center}
	\begin{figure}
	\centering
	\includegraphics[width=0.5\textwidth]{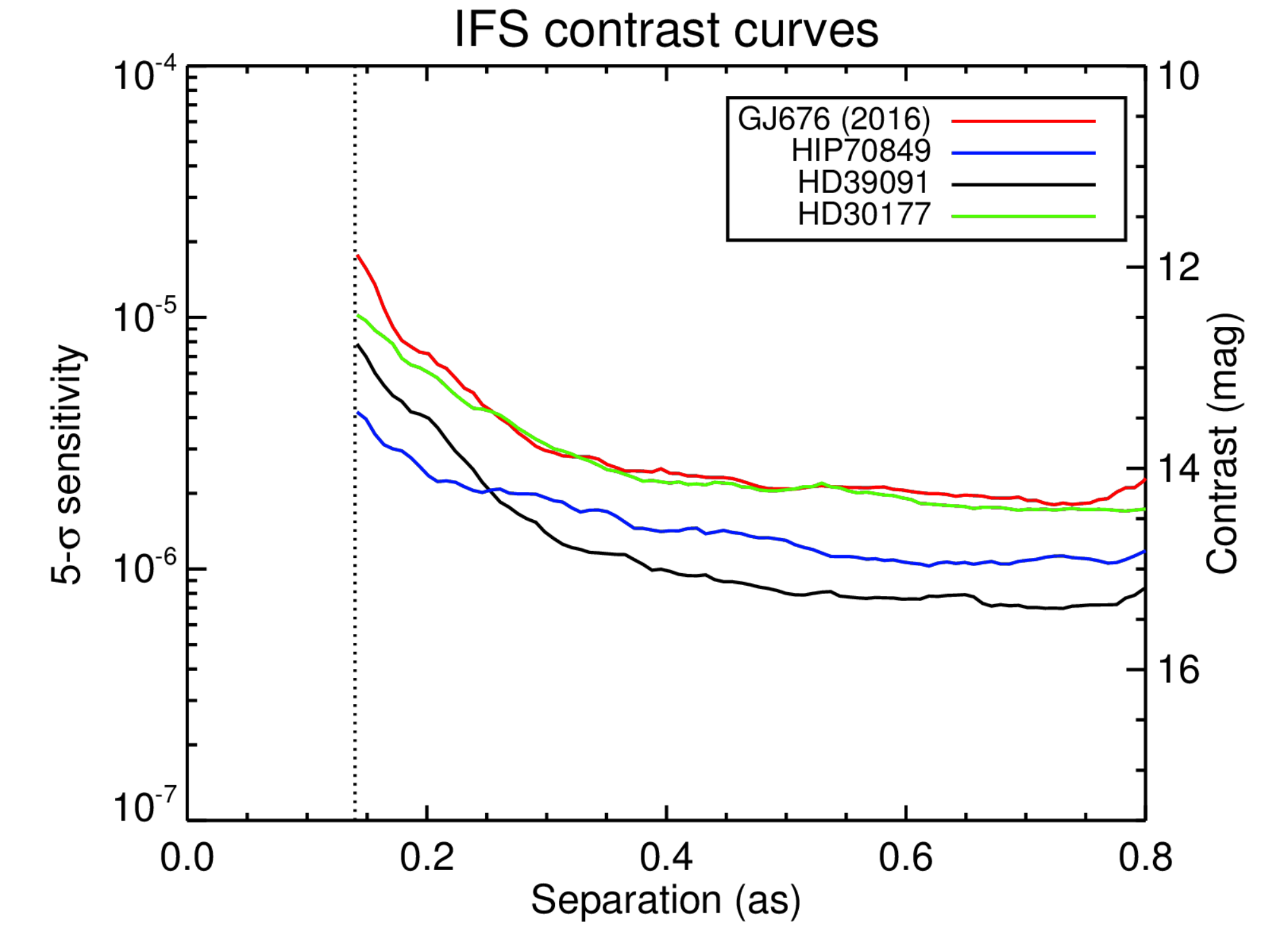}
	\caption{5$\sigma$ residual noise levels as a function of separation for the IFS data only. The curves refer to the observations of GJ\,676, HIP\,70849, HD\,39091, and HD\,30177.  Contrasts presented here were obtained with the PCA algorithm. }
	\label{f:con_rv}
        \end{figure}
\end{center}

\section{HD\,142}
\label{s:HD142}

The first target observed as part of this study, HD\,142, is in fact a double star system.  The primary, HD\,142\,A is a F7V star with a mass of $1.25 \pm 0.01$ $M_{\odot}$ \citep{2016A&A...585A...5B} and a distance to the Earth of 26.2 $\pm$ 0.1 pc
\citep{2018arXiv180409365G}.  The secondary, HD\,142\,B, is K8.5-M1.5 companion, with a smaller mass of 0.59 $\pm$ 0.02 $M_{\odot}$, at a projected separation of $\sim$4\as from the primary.
\par
The primary, HD\,142\,A, has two RV-detected companion objects: HD\,142\,Ab has a $M\sin{i}$ of 1.03~\MJup and a semimajor axis of 1.0 au, corresponding to $\sim$0\farcs04 \citep{tinney2002} while HD\,142\,Ac has a $M\sin{i}$ of 5.3~\MJup and a 
semimajor axis of 6.8 au, corresponding to $\sim$0\farcs26, with an eccentricity of 0.21 and a period of 6005 days \citep{wittenmyer2012}. 
For HD\,142\,Ac, our Monte Carlo simulations predicted a probability of detection of 11.30\% for the epoch of October 2014.
The orbital parameters used for the simulation are listed in Table~\ref{t:c}. However, we stress that, given the very low
minimum mass obtained from the RV measurement, we could realistically expect to be able to image HD\,142\,Ac only if the inclination of its orbit is quite far
from being edge-on. 

\begin{table*}
 \centering
 \begin{minipage}{140mm}
\caption[]{Orbital parameters for the planets of the systems presented in this analysis.}
\label{t:c}
\begin{tabular}{lcccccc}
\hline
\hline
Planet         & Period (days) & T$_{0}$ (JD-2,400,000)   & e &$\omega$ (deg)  &
$M\sin(i)$ (\MJup) & Ref. \\
\hline
HD\,142\,c &$ 6005 \pm 477$ &$55954 \pm 223$ &$ 0.21 \pm 0.07$ &$250 \pm20$ &$5.3 \pm 0.7 $ & \citet{wittenmyer2012} \\
GJ\,676\,c & 7462.9$^{+105.4}_{-101.4}$ &  405.4$^{+63.5}_{-65.6}$ & 0 & - & 6.9 $\pm$ 0.1 & \cite{2016sah}   \\
HIP\,70849\,b & 17349.4 $\pm$ 600 & - & 0.715 $\pm$ 0.245 & - & 9 $\pm$ 6 & \citet{Segransan11}    \\
HD\,39091\,b & 2049 $\pm$ 150 & 50073 $\pm$ 150 & 0.61 $\pm$ 0.03 & 330 $\pm$ 20 & 10.3 & \citet{2002MNRAS.333..871J} \\
HD\,30771\,c & 11613 $\pm$ 1837 & 51660 $\pm$ 573 & 0.22 $\pm$ 0.14 & 11 $\pm$ 13 & 7.6 $\pm$ 3.1 & \citet{2017AJ....153..167W} \\
\hline
\end{tabular}
\end{minipage}
\end{table*}


\subsection{System age}
\label{s:rot}

As the age of the system strongly impacts the interpretation of the observational data, we 
reconsidered the age estimate for the HD\,142 system here.
Activity and rotation age indicators provide contradictory age estimates for this system: 
the low chromospheric activity
\citep{2006MNRAS.372..163J,2014A&A...561A...7R} 
and non-detection in X-ray \citep{2010A&A...515A..98P} point to an age older than the Sun, when using 
\citet{2008ApJ...687.1264M} calibrations, however, the primary is a moderately fast rotator, with a rotational
velocity of $v \sin{i}=  10.4\pm0.4$\,kms$^{-1}$ \citep{2005ApJ...622.1102F,2006ApJ...646..505B}, 
implying a younger age. 
Combining the projected rotational velocity and the stellar radius (R$ = 1.44\pm0.14\,$R$_\odot$), 
we derive an upper limit for the stellar
rotation period P$_{\rm rot}$ $<$ 7 $\pm$ 1\,d.
This corresponds to an age of $< 1.3\pm0.3$\,Gyr, using the gyro-chronology relations 
of \citet{2008ApJ...687.1264M}.

On the basis of our derived upper limit on its rotation period and its F7V spectral type, we expect
that HD\,142A may exhibit magnetic activity and, therefore, photometric variability
arising from activity centers on its photosphere. The light rotational modulation induced by activity centers can
be used to derive the rotation period using Fourier analysis.
While this star has been included in a few photometric surveys, given its brightness
the only unsaturated photometry available is from Hipparcos, which obtained 214 photometric
measurements with a photometric precision $\sigma_{\rm V}$ = 0.006 mag between December 14, 1989 to March 17, 1993
We used the Lomb-Scargle \citep[LS;][]{1976Ap&SS..39..447L,1982ApJ...263..835S} and the CLEAN periodogram analysis techniques
to search for significant periodicities (due to the stellar rotation period) in the magnitude time-series. The false alarm probability (FAP = 1 $-$ confidence level) associated with our detected period, which is the probability that a peak of given height in the periodogram is caused simply by statistical variations, i.e., Gaussian noise, was computed through Monte Carlo simulations, i.e., by generating 1000 artificial light curves obtained from the real light curve, keeping the date but permuting the magnitude values (see, e.g., \citealt{Herbst02}).  We followed the method used by \cite{Lamm04} to compute the errors associated with the period determination.
No highly significant periodicities were detected for HD\,142A.

\newpage

The discrepancy between age estimates based on rotation and activity is likely due to the fact that 
the temperature / spectral type of this star is close to the blue edge of the range of valid spectral types / colors used 
in the \citet{2008ApJ...687.1264M} age-activity calibration (valid for F7-K2 dwarfs, with 0.5 mag $<~B-V~<$0.9 
and that its evolutionary phase is somewhat off of the main
sequence (see below). Indeed, the observed rotational velocity is qualitatively consistent with 
rotation evolution model from \citet{vansaders2013}, for a star matching the spectroscopic parameters of HD142\,A
from \citet{2014A&A...561A...7R}.
Lithium absorption features in the spectrum of this star do not provide a useful age constraint, as lithium 
is not particularly sensitive to stellar age for stars with temperatures similar to HD\,142\,A.
However, a young stellar age ($\le$ 1 Gyr) can be ruled out by the kinematic parameters for this star
\citep{2004A&A...418..989N}, well outside the space populated by young stars \citep{2001MNRAS.328...45M}.

The most reliable age estimate can be obtained through isochrone fitting.
From such an isochronal age analysis, 
\citet{2014A&A...561A...7R} recently derived an age of  $2.6_{-0.3}^{+0.2}$ Gyr. This estimate is compatible with the value 
of $2.8 \pm 0.5$ Gyr found by \citet{2016A&A...585A...5B}.
These estimates suggest an intermediate age for this system.  As these isochronal age estimates
are more accurate than the estimate based on other methods, we adopt an age of 2.6 Gyr for HD\,142 in the following analysis

\subsection{Results}
\label{s:dis}

From our data reduction, we do not detect any companions 
either at the expected separation of 0\farcs3 for planet c (blue curve in Fig.~\ref{f:hd142}), or at any other separations. 
This lack of detections can be seen in the ADI image in Figure~\ref{f:fi_hd142} for IRDIS and in the PCA image in
Figure~\ref{f:ifs_final} for IFS. \par

\begin{center}
	\begin{figure}
	\centering
	\includegraphics[width=0.5\textwidth]{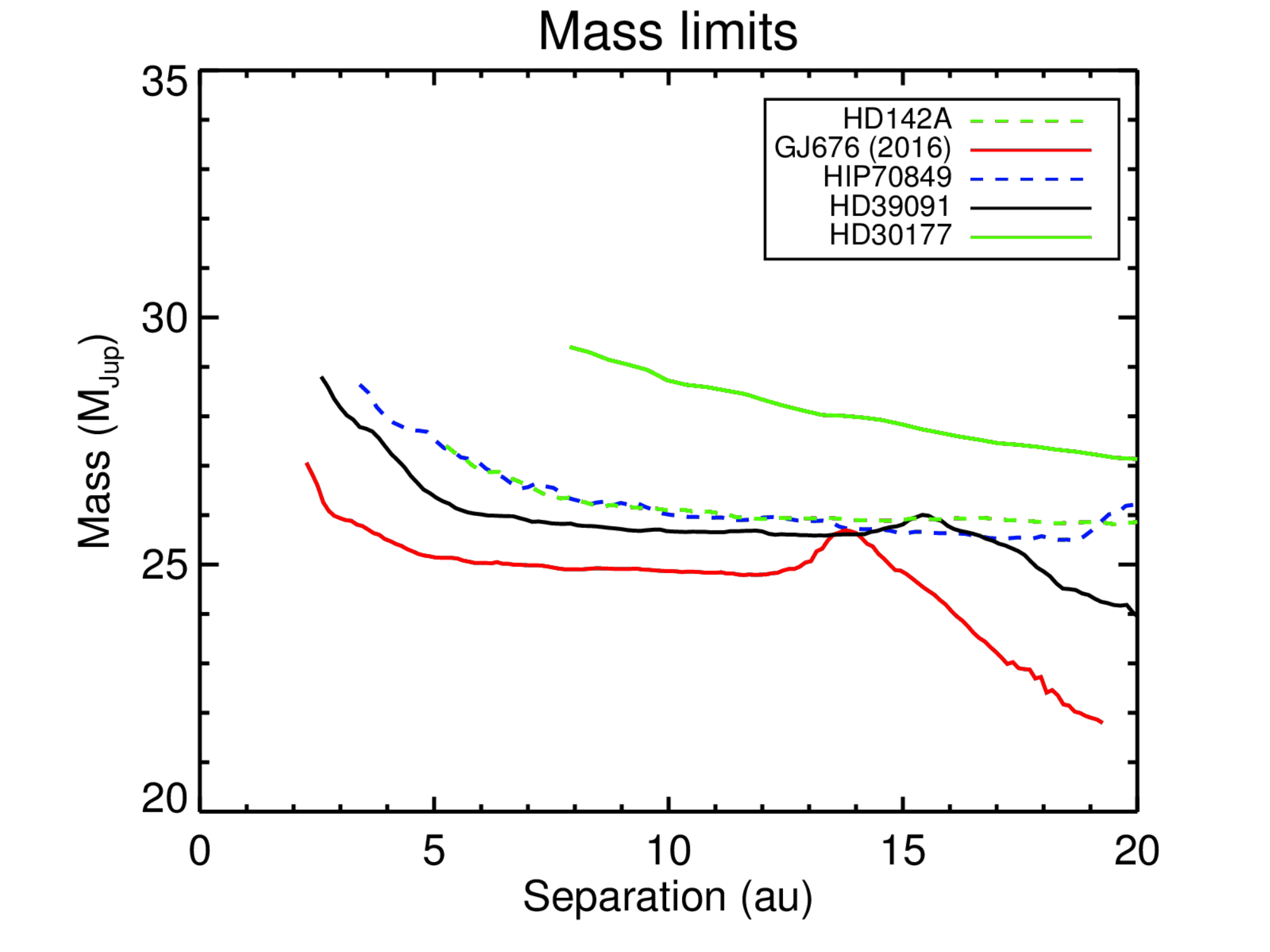}
	\caption{Mass limit versus the separation in au from the central star as obtained from converting the contrast curves
        displayed in Figures~\ref{f:con_hd142} and~\ref{f:con_rv} to minimum detectable companion mass using the AMES-COND models.}
 	
	\label{f:all_mass}
        \end{figure}
\end{center}

\begin{table}
 \centering
\caption[]{Summary table with the upper limit of the mass of planets around each of the five systems of this analysis at different separations from the star. The maximum possible inclination of each RV planet is also shown in the last column.}
\label{t:res}
\begin{tabular}{lcccc}
\hline
\hline
System & 5 au   &  10 au & 15 au & max(i) \\
\hline
HD\,142A& 28 \MJup & 26 \MJup &  26 \MJup &  11 deg\\
GJ\,676& 25 \MJup   & 25 \MJup & 25 \MJup & 16 deg \\
HIP\,70849& 28 \MJup & 26 \MJup & 26 \MJup & 20 deg \\
HD\,39091& 26 \MJup & 26 \MJup &  26 \MJup & 23 deg \\
HD\,30177& 30 \MJup & 29 \MJup &  28 \MJup & 15 deg \\
\hline
\end{tabular}
\end{table}

Contrast limits in flux as well as J band magnitude are shown in Figure~\ref{f:con_hd142}, after correcting for self-subtraction via the procedure described 
in Section~\ref{s:obs}.   Adopting a stellar age of 
2.6 Gyr, as justified in Section~\ref{s:rot}, 
we calculated the mass limits displayed in Figure~\ref{f:all_mass}, and listed in Table~\ref{t:res}, by converting our contrast limits to minimum detectable 
companion mass using the AMES-COND 
models \citep{2000ASPC..212..127A}. As previously mentioned, we used for this work only the contrast limits obtained with IFS,
since the contrast we obtain with this subsystem is deeper than that gathered with IRDIS. At the expected separation for HD\,142\,Ac of 0\farcs3, 
we obtained an upper mass limit of $22.6_{-1.7}^{+0.9}$~\MJup for a T-type spectrum and $23.6_{-1.7}^{+0.9}$~\MJup for an L-type spectrum.
Using the AMES-COND models we can infer an upper limit for T$_{eff}$ of $\sim$ 600 K for HD\,142\,Ac from the mass limit calculated
in this work, while using the minimum mass obtained through the RV we can infer a minimum value for the T$_{eff}$ of $\sim$ 270 K.

Our upper mass limit for HD\,142\,Ac constrains the possible inclination of its orbit, 
excluding all inclinations smaller than $13.6_{-0.5}^{+1.1}$ deg from a pole-on orbit. Given the 
minimum mass of 5.3~\MJup for this planet obtained through the RV technique and the upper 
mass limit obtained with SPHERE, the orbit of HD\,142,\,Ac would have had to have been very close
to the pole-on case in order for SPHERE to have been able to image it. 

\par

The MESS \citep[Multi-purpose Exoplanet Simulation System, see][]{2012A&A...537A..67B, 2013PASP..125..849B} code was used to evaluate the probability of detection of companions around HD 142 A. 
The code is a recently developed tool with a well demonstrated utility for analysis of exoplanet data \citep[see e.g][]{2015A&A...573A.127C, 2013A&A...553A..60R}

\subsection{HD\,142\,B}
\label{s:comp}

\begin{center}
	\begin{figure}
	\centering
	\includegraphics[width=0.5\textwidth]{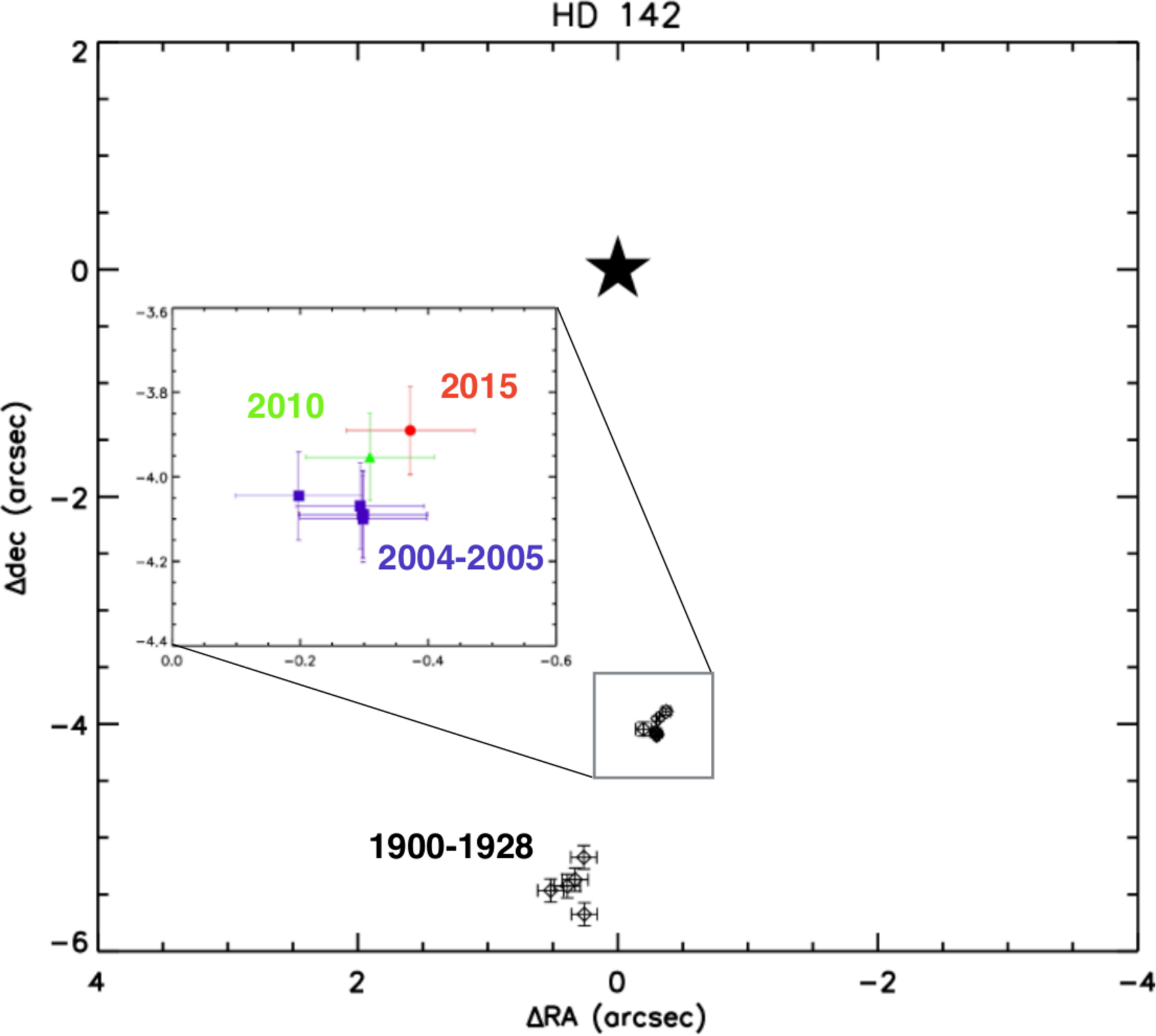}
	\caption{Astrometric positions of HD\,142\,B with respect to HD\,142\,A obtained from measurements starting from 1894. The error bars
        for each measurements are shown. More details are found in Table~\ref{t:age}. The black star at the center of the Figure is the position of HD\,142\,A.}
 	
	\label{f:astropos}
        \end{figure}
	\end{center}

HD\,142\,A has a stellar companion, a K8.5-M1.5 star with a mass of $0.59 \pm 0.02$~M$_{\odot}$ \citep{2007A&A...474..273E}. 
Combining our SPHERE astrometry with astrometry from the literature, we can place new constraints on the orbital elements for this stellar companion.
Since HD\,142\,B is strongly saturated in our coronagraphic images, 
we measured its position and contrast using non-coronagraphic IRDIS images which are normally taken for flux calibration purposes. 
For these flux calibration images, a neutral density filter is used, preventing saturation for both stars in this binary system. For astrometric calibrations (true North, platescale, distortion), we refer the reader to \citet{2016SPIE.9908E..34M}. The astrometric and photometric results for this object are shown in Table~\ref{t:age}, together with the values available in the literature.
We retrieved all the relative astrometric measurements available in the Washington Double Star Catalog \citep[WDS;][]{2001AJ....122.3466M}, starting from 1894. To these data we added the results of \citet{2007A&A...474..273E} and \citet{wittenmyer2012}, as well as our own measurement, which are listed in Table~\ref{t:age}. All these astrometric positions are displayed in Figure~\ref{f:astropos} together with the error bars that we adopted to use with our Monte Carlo simulation of potential orbital parameters. As no error bar was given for the WDS point, we assumed a reasonable value for the given instrumentation of that epoch.  

\begin{table*}
\centering
\begin{minipage}{140mm} 
\caption[]{Astrometric positions of the stellar companion HD\,142\,B.}
\label{t:age}
\begin{tabular}{lcccccc}
\hline
\hline
Epoch         &PA (deg) & $\sigma_{PA}$ (deg)   &  $\rho$ (\as)   &$\sigma_{\rho}$ (\as)  & $\Delta$mag &  Ref. \\
\hline

       1900.01  &   177.1  &   -    &    5.18   &   - & -  &1\\
        1900.86  &   175.9  &   -    &    5.44  &    -& - &1\\
        1901.90   &  177.4   &  -   &     5.68   &   -& -&1\\
        1914.77  &   174.6  &   -   &     5.49   &   - & -&1\\
        1928.85  &   176.5  &   -    &    5.38   &  -& -&1\\
       2004.4762  & 184.16 &  0.18  &   4.10   &   0.02& $\Delta H =3.09 \pm 0.05$  &2     \\   
       2004.4788  & 184.16 &  0.18  &   4.11   &   0.02& $\Delta H =3.52\pm 0.07$  &2    \\    
       2004.8505  & 184.18 &  0.18  &   4.10  &    0.02& $\Delta H =3.09\pm 0.04$  &2    \\    
       2005.9373  & 184.13 &  0.29  &   4.08  &    0.02 & $\Delta H =2.89 \pm 0.15$ &2   \\ 
       2010.5377  & 184.47 &  0.26  &   3.965  &   0.013 &- &3\\ 
       2014.7847 & 185.462 & 0.04  &   3.908  &   0.002& $\Delta H2 = 3.39 \pm  0.04$   & 4\\ 
                  & & & & &$\Delta H3 = 3.08 \pm 0.08$ &4 \\
\hline

\end{tabular}
\tablebib{
  1: \citet[WDS;][]{2001AJ....122.3466M};
2: \citet{2007A&A...474..273E};
 3: \citet{wittenmyer2012};
 4: This work}
\end{minipage}
\end{table*}

Using all the available astrometric points, we performed a Monte Carlo simulation of potential orbits, following the model by \citet{2013A&A...554A..21Z} and \citet{2011A&A...533A..90D}. From the results of this simulation, we excluded the orbits that may cause instability in the system, following Eq.~1 of \citet{1999AJ....117..621H}. Orbits where the critical semi-major axis is greater than the periastron of planet c were thus excluded from the results. Although the equation~1 in \citet{1999AJ....117..621H} refers to circular and coplanar orbits, it is a good approximation to exclude solutions where the eccentricity is too high to assure the stability of the system. 

\begin{center}
	\begin{figure*}
	\centering
	\includegraphics[width=\textwidth]{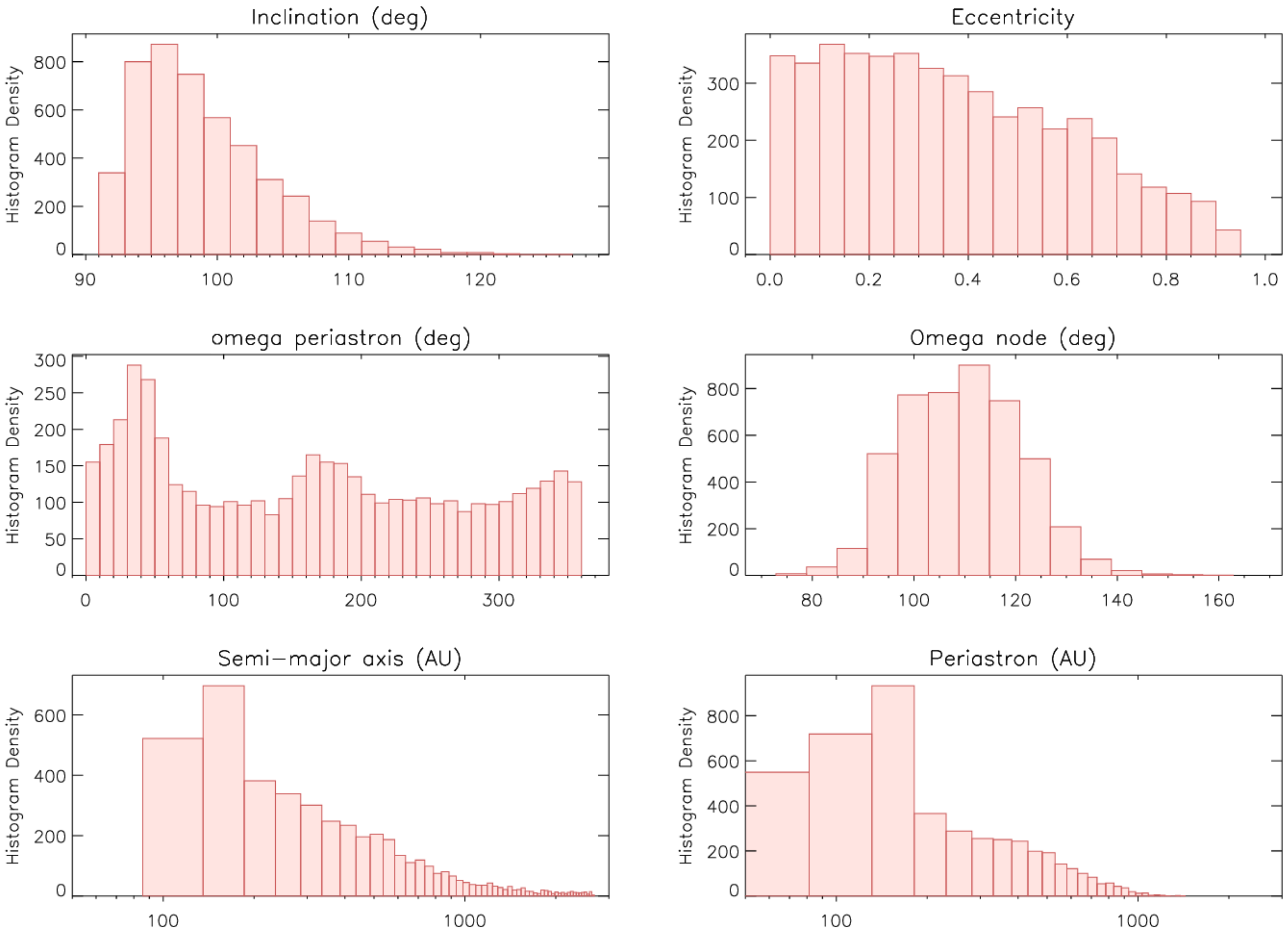}
	\caption{Histograms with the distribution of the orbital parameters of HD\,142\,B from our Monte Carlo simulation. The orbital parameters do not follow uniform distributions, but do present clear maximum probability values. Orbits which produce unstable systems have been excluded as explained in the text.  }
 	
	\label{f:histo}
        \end{figure*}\end{center}

From this Monte Carlo simulation, we constrain the orbital parameters of the stellar companion.  The histograms of the resulting orbital parameter distributions are plotted in Fig.~\ref{f:histo}.
The orbit of HD\,142\,B is most likely nearly edge-on (peak at 96 deg) and the semi-major axis is most likely around a = 150 au.  A caveat to mention is that as we used a uniform linear distribution to generate the semi-major axis values, we might be biased towards long period orbits. Considering the limits placed on system inclination from the non-detection of HD\,142\,Ac and from the measured orbit of the stellar companion HD\,142\,B, the orbits of HD\,142\,Ac and HD\,142\,B are compatible with coplanarity but substantial misalignments are also possible. If the coplanarity of HD\,142\,B and HD\,142\,Ac were confirmed, the mass of the latter would be very near to its minimum mass of $\sim$5~\MJup, beyond the limits of detection with current direct imaging instrumentation.

\section{GJ\,676}
\label{s:GJ}

GJ\,676\,Ab \citep{2011A&A...526A.141F} is a sub-stellar companion orbiting the binary star Gliese\,676\,A. The system is composed of two M stars with an angular separation of about 50 arcsec at a distance of 16.0 $\pm$ 0.1 pc \citep{2018arXiv180409365G}; GJ\,676\,A has been classified as a M0V star \citep{2010MNRAS.403.1949K}.

As with HD\,142, we reconsider the age estimate for GJ\,676\,A as well.
Spectroscopic time series data collected by \cite{Suarez-Mascareno15} for this star revealed a rotation period P = 41.2$\pm$3.8\,d.  
\cite{Suarez-Mascareno15} measured magnetic activity level for GJ\,676\,A  via the calcium R$^{\prime}_{\rm HK}$ index, obtaining a value of Log R$^{\prime}_{\rm HK}$ = -4.96$\pm$0.04. The presence of activity cycles were reported by \cite{Suarez-Mascareno16} with a period P = 7.5$\pm$0.6\,yr and an amplitude of 0.08\,mag from photometric time series.  \cite{Gomes-da-Silva12} also report activity cycles with a P = 3.35\,yr based on NaI EW time series.\\
Using the \cite{Mamajek08} gyro-age relation, we inferred an age of 3.6$\pm$0.5 Gyr, based on the known rotation period and the stellar color of B$-$V = 1.44\,mag (\citealt{Koen10}). 
We derived a similarly old age 4.6$\pm$0.4\,Gyr from the gyro-age relation of \cite{Engle11}, which is valid for M0V stars.
Based on the known level of chromospheric activity, we can also infer a chromospheric age of  about 5.8\,Gyr using the activity-age relation by \cite{Mamajek08}. Despite the divergence of age estimates drawn from different methods, it is clear that GJ\,676 has an age equal to or older than about 3.6\,Gyr. 

GJ\,676\,Ab has a minimum mass of  4.9~\MJup and a period of 1056.8~days.  The planet is on a low eccentricity orbit, with no constraints on inclination from astrometry.  
Based on the Monte Carlo simulations described in Section~\ref{s.sel}, the minimum companion mass detectable by SPHERE is 41~\MJup and the probability of detecting GJ\,676\,Ab is 10\%.  \citet{2012A&A...548A..58A} rederived Doppler measurements in the HARPS South database for this system and discovered a number of additional candidate companions; GJ\,676\,A may hosts up to four sub-stellar companions.   Of these companions, GJ\,676\,Ac, which shows a trend in the RV, may also observable with SPHERE, with a minimum mass of 19.3~\MJup. 
\citet{2011A&A...526A.141F} provide orbital parameters for GJ\,676\,Ab, but the uncertainties on these values is quite high.  Recently, \citet{2016sah}, constrained the mass of planet b using astrometry. These authors observed the primary star with FORS2 at VLT to obtain precise astrometric positions over a period of 2 years starting in April 2010 and determined an orbit for planet b of 1052 days based in the motion of the star around the center of mass of the system. From this value they derived a mass for this planet of 6.7$^{+1.8}_{-1.5}$ \MJup. This value is compatible with our upper mass limit of 27 \MJup for the planet based on our SPHERE observations.  For planet c, we find an upper mass limit of 25 \MJup, but since the two objects are likely coplanar the expected mass for planet c would be close to its minimum mass of 6.8 \MJup -- thus, it is too small and faint to be imaged with current instruments, {as the contrast needed would be of the order of almost 40 magnitudes}.  

\section{HD\,39091}
\label{s:HD39}
HD\,39091\,b \citep{2002MNRAS.333..871J} is a massive object with a minimum mass of 10.09~\MJup, long period (2151~days), and very eccentric (0.641) orbit. It orbits a G1 star at a distance of 18.28 $\pm$ 0.2~pc from the Earth \citep{2018arXiv180409365G}.  \citep{2011A&A...527A.140R} provide constraints on the astrometry of this system which limit the inclination of this object to 20-150 $\deg$.  Unfortunately, with this range of inclinations, this object is undetectable with SPHERE.   The existing astrometric constraints are directly taken into account in the simulation, when generating all the possible random inclinations. For this target we reached a contrast of 10$^{-6}$ at 0\farcs3-0\farcs8 (see Figure~\ref{f:con_rv}).  From this contrast limit, the companion must be more massive than 26 \MJup at separations wider than 5 au as shown in Figure~\ref{f:all_mass} and listed in Table~\ref{t:res}.

\cite{Bonfanti16}  estimate an age of 2.8$\pm$0.8\,Gyr for HD 39091 using the isochronal fitting method. \cite{Pace13} estimate a chromospheric age of 4.03$\pm$1.33\,Gyr, while \cite{Saffe05} derive 
chromospheric ages of 3.83\,Gyr and 1.83\,Gyr, respectively, depending on the adopted calibration (either \cite{Donahue93} or \cite{Rocha-Pinto98}).\\
We used archival time series photometry to estimate a gyrochronological age for this system.
As this star is relatively bright, the only time series photometry of HD\,39091 available is from the Hipparcos archive (ESA 1997). We retrieved a total of 127 measurements collected between November 1989 and March 1993, with an average photometric precision $\sigma_V$ = 0.006\,mag.  We inferred a stellar radius R = 1.14$_{\odot}$  from the visual magnitude V = 5.67\,mag (SIMBAD database) for this star.  Combining this stellar radius with the distance d= 18.28 pc (the bolometric correction BC$_{\rm V}$ = $-$0.06\,mag \citep{Pecaut13} appropriate for its G0V spectral type, and the average projected rotational velocity v$\sin{i}$ = 2.96\,km\,s$^{-1}$ (\citealt{Delgado15}), we expect a rotation period P $\la$ 20\,d.  However, for the possible rotational light modulation to be detectable in the Hipparcos time series, the inclination of the  stellar rotation axis must be $i$ $\ga$ 20-30$^{\circ}$.  If this was the case, we would expect a much shorter rotation period P $\ge$ 5\,d.  Thus, we carried out our  LS and CLEAN periodogram analyses in the period range from 5 to 20 days and detected a period of P = 18.3$\pm$1.0\,d with a 99\% confidence level, which can be interpreted a the stellar rotation period.  We fit a sinusoid to the time series photometry phased by the rotation period, finding a peak-to-peak amplitude for the best-fit sinusoid of $\Delta$V = 0.008\,mag.  Using the \cite{Mamajek08} gyro-age relation, we inferred from the known rotation period and the stellar color B$-$V = 0.58\,mag (SIMBAD database) an age of 3.4$\pm$0.4 Gyr. 
Combining results from these three different age determination methods, we estimate the age of HD\,39091 to be $\sim$3.5\,Gyr.

\begin{figure*}
\begin{minipage}{18cm}
\includegraphics[scale = 0.6, trim = 0 0 0 70, clip, angle=90]{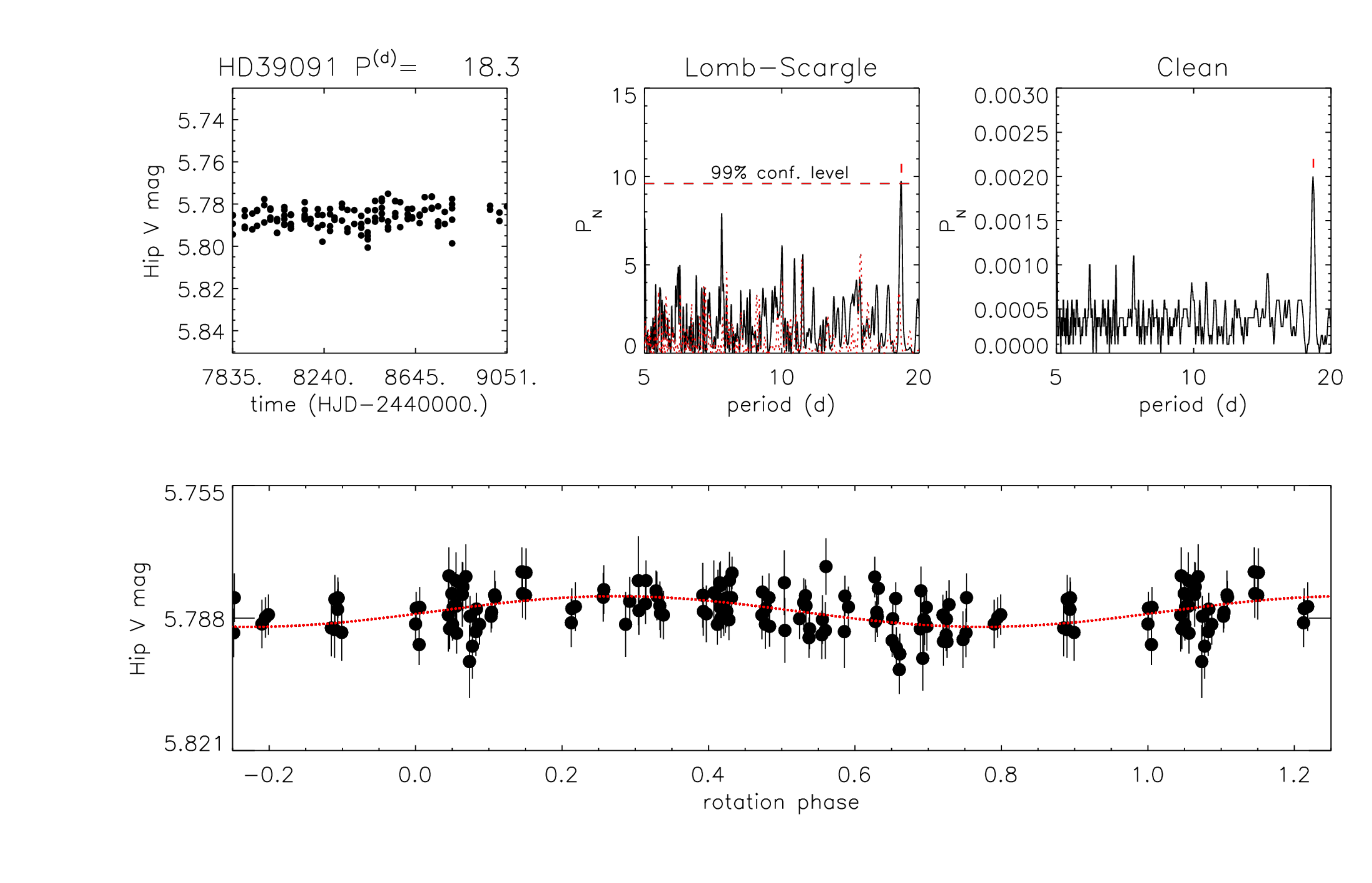}
\end{minipage}
\caption{\label{HD30177} \it  Top panel: \rm (left) ASAS V-magnitude time series for HD\,39091 versus Julian Day; (center) Lomb-Scargle periodogram (black) with the window function over-plotted (dotted red) and the power level corresponding to a 99\% confidence level. The vertical line marks the power peak corresponding to the stellar rotation period; (right) CLEAN periodogram.  \it  Bottom panel: \rm ASAS V-magnitude plotted versus rotation phase, computed using the period P = 18.3\,d. The red solid line is a fit to the data with a sinusoidal function.}
\end{figure*}

\section{HIP\,70849}
\label{s:HIP}
HIP\,70849 is an old (1-5 Gyr) K7V star, with a mass of 0.63 $\pm$ 0.03 \MSun \citep{2011A&A...535A..54S} and a distance of 24.1 $\pm$ 0.1 pc \citep{2018arXiv180409365G} from the Earth.   \citet{2011A&A...535A..54S} detected HIP\,70849b, a long-period (5-90 yr), high eccentricity planet orbiting this star, with a mass between 3-15 \MJup.  
\citet{2014A&A...569A.120L} identified as well an extremely wide T4.5 dwarf companion at a distance of 6.3 arcmin, or $\sim$ 9100 au, well outside of the FOV of the IRDIS detector.

HIP\,70849 exhibits clear evidence of magnetic activity from both photometric and chromospheric proxies. 
HIP70849 was observed by SuperWASP from May 2006 until April 2008. A total of 10160 measurements were collected. The V-band magnitude time series exhibits a decreasing linear trend that was removed before our period search. After outlier removal and data binning (24-hr bin width), we ran the Lomb-Scargle and CLEAN periodograms on a total of 123 mean magnitudes and detected a significant (FAP $<$ 0.01) power peak at P = 41.2$\pm$0.4\,d in both periodograms. The sinusoidal fit to the light curve has an amplitude of $\Delta$V = 0.007\,mag. 
The 41.2-d periodic light modulation is likely the stellar rotation period and arises from the presence of surface brightness inhomegeneities that are carried in and out of view by the stellar rotation.
 This star also shows a starspot cycle of  P$_{cyc}$ = 10.1$\pm$1.4 yr with an amplitude of A$_{cyc}$ = 6.9$\pm$0.9\,mmag \citep{SuarezMascareno16}.\\

To derive the basic physical parameters of this star we analyzed the spectral energy distribution (SED). 
We used the VOSA tool \citep{Bayo08} to build the observed SED, which was best fitted by a model spectrum from the BT-NextGen-GNS93 suite of models\citep{Allard12} with T$_{eff}$ = 4000$\pm$50\,K, gravity log g = 4.5 and metallicity [Fe/H] = 0.3. We additionally inferred a bolometric luminosity L$_{bol}$ = 0.0892$\pm$0.0005\,L$_{\odot}$, and a stellar radius R = 0.62$\pm$0.02\,R$_{\odot}$.
Rotation period, stellar radius, and the projected rotational velocity $v\sin{i}$ = 1.93\,km\,s$^{-1}$ \cite{Segransan11} can provide an estimate of the rotation axis inclination. However, we inferred an inconsistently large  $\sin{i}$ = 2.7, likely arising from an overestimated  $v\sin{i}$ value. 
In fact, the combined effect on spectral lines by the macro- and micro-turbulence  in main sequence  K7-type stars (\citealt{Gray84}; \citealt{Husser13}), although relatively small ($<$ 1 km\,s$^{-1}$), is enough to explain the overestimated rotational velocity.  Taking this into account, we infer an inclination $i$ $\sim$ 90$^{\circ}$. 

\cite{SuarezMascareno16} used All Sky Automated Survey (ASAS) photometry to investigate the rotational and magnetic activity properties of this target. They detect a starspot cycle of  P$_{cyc}$ = 10.1$\pm$1.4 yr with an amplitude of A$_{cyc}$ = 6.9$\pm$0.9\,mmag.
However, as this period is longer than the 8.8-yr time span of the ASAS timeseries data, it should be regarded as tentative. They also marginally detect periodic modulation of the light curve with an initial period estimate of  P$_{rot}$ = 133.55$\pm$0.75\,d.  This star has chromospheric emission with an index R$^{\prime}_{HK}$ = -4.74$\pm$0.05 and a projected rotational velocity $v\sin{i}$ = 1.93\,km\,s$^{-1}$ measured from the HARPS cross-correlation function.

Using the age-rotation relationship from \citet{Mamajek&Hillenbrand08},  we derive a gyrochronological age of 3.6$\pm$0.15 Gyr
from the P = 41.2\,d rotation period.   Using the age-activity relationship from \citet{Mamajek08}, we derive a chromospheric  age of 2.43\,Gyr from the activity index R$^{\prime}_{HK}$ = -4.74$\pm$0.05.
Finally, the relatively long starspot cycle collocate this star in the active branch of the rotation period-rossby number relation, which is populated mainly by old field stars \citep{1999ApJ...524..295S}.

We found a detection probability of $\sim$20\% for HIP\,70849b using our Monte Carlo simulation, however, no companion was detected with IFS or IRDIS in our SPHERE observations.   Our observations obtained a mean contrast of 2$\times$10$^{-6}$ at 0\farcs3-0\farcs8 (see Figure~\ref{f:con_rv}).  Adopting a relatively old age for the system (as discussed above), the companion must be more massive than 26 \MJup at separations wider than 7 au as shown in Figure~\ref{f:all_mass} and summarized in Table~\ref{t:res}.

\section{HD\,30177}
\label{s:HD30}

\citet{2017AJ....153..167W} detected a candidate massive Saturn-analog orbiting the solar-type star HD\,30177.  The system was previously known to host an interior planet with a minimum mass of 9.7 \MJup ; 
the outer planet has msini = 7.6 $\pm$ 3.1 \MJup and a = 9.9 $\pm$ 1.0 au.  

Isochronal and chromospheric age estimates of HD\,30177 suggest this star is rather old. 
\cite{Bonfanti16} and \cite{Ramirez12} estimated ages of 5.9$\pm$1.1\,Gyr and 6.18$\pm$2 respectively via isochronal fitting, while \cite{Saffe05} found a chromospheric age for the system 
of 8.30\,Gyr or 1.50\,Gyr, depending on the adopted calibration (\cite{Donahue93} or \cite{Rocha-Pinto98} respectively).
We used archival time series photometry to estimate a gyrochronological age as well for this system from the rotation period of the star.
The stellar rotation period in late-type stars, as HD\,30177, can be inferred from the period of the light rotational modulation arising from the presence of surface temperature inhomogeneities related to the magnetic activity.
HD\,30177 was observed by the ASAS survey \citep[All Sky Automated Survey][]{Pojmanski97} from 2000 to 2009. 
After removing outliers with a 3-$\sigma$ moving boxcar filter and discarding a few inaccurate ($\sigma_{\rm V}$ $\ge$ 0.05 \,mag) data points, a time series of 510 measurements remained, characterized by an average photometric precision $\sigma_{\rm V}$ = 0.03\,mag.  This time series was analysed with the Lomb-Scargle (LS; \citealt{Scargle82}, \cite{Horne86})  and the CLEAN (\citealt{Roberts87}) periodogram methods to search for significant periodicities.\\

HD 30177 has an estimated radius R = 1.16$\pm$0.31\,R$_{\odot}$. This value is derived from the visual magnitude V = 8.37\,mag \citep{Hog00}, the distance d = 55.7 $\pm$ 0.1\,pc \citep{2018arXiv180409365G}, and the bolometric correction BC$_{\rm V}$ = $-$0.14 \citep{Pecaut13} corresponding to an average effective temperature T$_{\rm eff}$ = 5596\,K (see Table 1 in \citealt{Wittenmyer17}).
Combining the estimated radius and the projected rotational velocity v$\sin{i}$ = 2.96$\pm$0.5\,km\,s$^{-1}$ \citep{Butler06}, the stellar rotation period is expected to be shorter than about 30\,d. On the other hand, for the possible light rotational modulation induced by surface inhomogeneities to be detectable, the stellar rotation axis of HD\,30177 must have an inclination sufficiently far from a pole-on configuration (i.e. ${i}$ $\ge$ 20-30$^{\circ}$).
These circumstances made reasonable a search from periodicities in the photometric time series of HD\,30177, covering a range of possible rotation periods from 10 to 30 days.\\

The LS periodogram analysis detected  a period  P = 24.9$\pm$1\,d, which can be likely interpreted as the stellar rotation period, although only with a confidence level $\leq$90\%. However, the same period is also found by the CLEAN algorithm, rendering this period more believable.  \\
The peak-to-peak amplitude of the sinusoid function used to fit the magnitudes phased with the rotation period is $\Delta$V = 0.007\,mag.
We note, for instance, that the very low amplitude of the photometric variability is consistent with the slow rotation of HD\,30177, whose level of activity is expected consequently low. Combining the stellar radius, the rotation period, and the projected rotational velocity we derive an inclination of the stellar rotation axis close to ${i}$ $\simeq$ 90$^{\circ}$. This circumstance likely allowed us to detect the rotational light variation, despite the very low level of magnetic activity and the relatively small ratio of the light curve amplitude over the photometric precision.\\

Using the \cite{Mamajek08} gyro-age relation, we inferred an age of 2.8$\pm$0.3 Gyr from the known rotation period and the stellar color B$-$V = 0.81\,mag (\citealt{Hog00}). This result is in qualitative agreement with the earlier results from the literature and points toward an old age for HD\,30177.

From the contrast limit that we obtained with SPHERE, the companion must be more massive than 30 \MJup at separations wider than 5 au as shown in Figure~\ref{f:all_mass} and listed in Table~\ref{t:res}.

\begin{figure*}
\begin{minipage}{18cm}
\includegraphics[scale = 0.6, trim = 0 0 0 70, clip, angle=90]{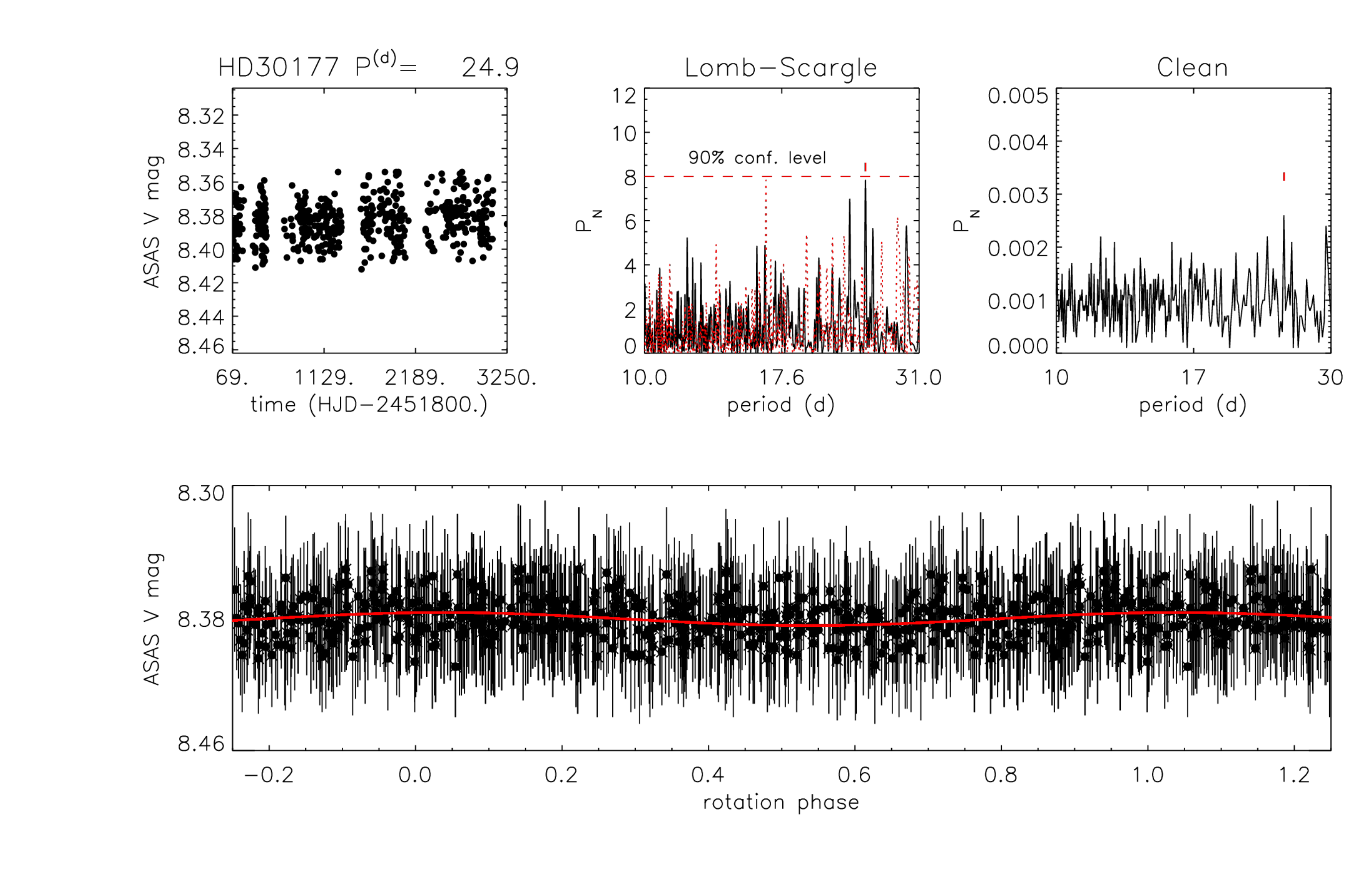}
\end{minipage}
\caption{ As in Fig.~\ref{HD30177}, but for the target HD\,30177. }
\end{figure*}

\section{Conclusions}
\label{conclusion}

We present in this paper the analysis of a sample of radial velocity systems imaged with SPHERE.  We selected a sample of 5 promising RV-detected, massive, wide orbit companions using a Monte Carlo simulation that explored all the possible inclinations for these objects.   These five sample stars, specifically, HD\,142\,Ac, GJ\,676\,b and c, HD\,39091\,b, HIP\,70849\,b, and  HD\,30177\,c, were observed during GTO using the NIR arm of SPHERE, reaching contrasts of 1 to 3 $\times$ 10$^{-6}$ {at separations larger than 0\farcs3}. We detected none of the known radial velocity companions for these stars, however, for the system HD\,142\,AB, the known stellar companion was detected in the IRDIS FoV.

We thus obtained an additional high precision astrometric point for HD\,142\,B, which, when combined with literature astrometry, provides a baseline of 130 years of observations. We used a Monte Carlo simulation to investigate the probability distribution of the parameters of its orbit, finding that the stellar companion most likely has an edge-on orbit with a semi-major axis of about 150 au. The orbit of HD\,142\,Ac might be coplanar with that of 
HD\,142\,B.  If this is the case, the mass of HD\,142,\,Ac may be very near to the minimum mass obtained through the RV measurements for this planet. For this reason, and given 
the mass limit found in this work, we can conclude that HD\,142\,Ac is considerably too low mass and too faint to be imaged with SPHERE.

From the contrast limits reached by SPHERE for the four other systems observed, we similarly conclude that the inclinations of these orbits are quite close to edge-on and the companion masses are close to the minimum mass given by RV measurements.

In the next few years, Gaia will provide astrometric measurements with precisions down to a few tens of microarcseconds.  Gaia astrometry will strongly constrain the inclination of the orbits of these and other similar planets and will also yield numerous new planet discoveries.  Given the performance of the current suite of planet imagers, detecting cool and old RV companions is highly challenging, however, with future 30m class telescopes, these objects will be easily detected and characterized. This will eventually yield vital constraints on evolutionary models for planetary mass companions.

\section*{Acknowledgments}

We want to thank the anonymous referee for his/her comments that improved the quality of this publication. We are grateful to the SPHERE team and all the people at Paranal for the great effort during SPHERE early-GTO run.  A.Z. acknowledges support from the CONICYT + PAI/ Convocatoria nacional subvenci\'on a la instalaci\'on en la academia, convocatoria 2017 + Folio PAI77170087. S.D., R.G. acknowledge support from the ``Progetti Premiali'' funding scheme of the Italian Ministry of Education, University, and Research. We acknowledge support from the 
French National Research Agency (ANR) through the GUEPARD project grant ANR10-BLANC0504-01. \par
SPHERE was funded by ESO, with
additional contributions from CNRS (France), MPIA (Germany), INAF (Italy),
FINES (Switzerland) and NOVA (Netherlands). SPHERE also received funding
from the European Commission Sixth and Seventh Framework Programmes as
part of the Optical Infrared Coordination Network for Astronomy (OPTICON)
under grant number RII3-Ct-2004-001566 for FP6 (2004-2008), grant number
226604 for FP7 (2009-2012) and grant number 312430 for FP7 (2013-2016).\par
This research has made use of the Washington Double Star Catalog maintained at the U.S. Naval Observatory. 
We thank Dr. B. Mason for making the individual measurements of HD142 available to us. \par

\bibliographystyle{mnras}
\bibliography{rv_targets}

\appendix
\section{Probability of detection}
\label{a:prob}
Concerning the probability of detection, it is clear that massive objects, for a given age, have a higher chance of being detected. We do not know the true inclination for a given planet, but we can calculate 
the probability that the planet mass is greater than the minimum detectable mass by integrating over a reasonable distribution of potential inclination values. 
For this aim, we used the {\it a posteriori} probability distribution of $sin(i)$ suggested by \citet{2011ApJ...739...26H}.
This inclination distribution \textit{a priori} is isotropic with the constraints given by \citet{2011A&A...527A.140R}.

The \textit{a priori} probability distribution function of the observed mass M$_0$ given the real mass M$_T$ is:
  	\begin{equation}
	P(M_0 \mid M_T) = \frac{M_0/M_T^2}{\sqrt{1-(\frac{M_0}{M_T})^2}}.
	\end{equation}
If we want the \textit{posterior} probability, using Bayes' theorem:
  	\begin{equation}
	P(A \mid B) = \frac{P(B \mid A)P(A)}{P(B)},
	\end{equation}
and assuming that
  	\begin{equation}
	P(M_0) = \int{P(M_0 \mid M_T)P(M_T)}dM_T,
	\end{equation}
we finally obtain
  	\begin{equation}
	P(M_T \mid M_0) = \frac{P(M_0 \mid M_T)P(M_T)}{\int{P(M_0 \mid M_T)P(M_T)}dM_T},
	\end{equation}
that is the \textit{posterior} distribution of the probability. We can write it as:
  	\begin{equation}
	P(M_T \mid M_0) = \frac{\frac{M_0/M_T^2}{\sqrt{1-(\frac{M_0}{M_T})^2}}P(M_T)}{\int{\frac{M_0/M_T^2}{\sqrt{1-(\frac{M_0}{M_T})^2}}P(M_T)}dM_T}.
	\end{equation}
If we want the probability of finding a mass greater than a value $X$:
  	\begin{equation}
	P(M_T > X \mid M_0) = \frac{\int_{M_0}^X{\frac{M_0/M_T^2}{\sqrt{1-(\frac{M_0}{M_T})^2}}P(M_T)dM_T}}{\int_{M_0}^{M_{max}}{\frac{M_0/M_T^2}{\sqrt{1-(\frac{M_0}{M_T})^2}}P(M_T)}dM_T}.
	\end{equation}
Finally, to calculate the probability we need the distribution of the true mass $P(M_T)$. In this work we assumed that the distribution in the planetary regime is a power law function with index $-1$ \citep{2010ApJ...714.1570H} and for the brown dwarf regime a power law with index approximately zero \citep{2009ApJS..181...62M}.

So, the function $P(M_T)$ is $\propto M^{-1}$ from $M_0$ to the upper limit of the brown dwarf desert and constant for greater values of the mass. The final function is the sum of the two different curves connected in the point where the mass is equal to the value of the upper limit of the brown dwarf desert. 
The boundaries of the brown dwarf desert are assumed to be 20-57~\MJup \citep{2006ApJ...640.1051G}.

If we want to know how many object we can detect with SPHERE in the substellar regime the upper limit of the mass is 80~\MJup. If the physical maximum mass of the target is in the planetary regime we changed the upper limit of the integral. 

In general we have to solve the integral:
  	\begin{equation}
	y = \int_{M_0}^X{\frac{M_0/M_T^2}{\sqrt{1-(\frac{M_0}{M_T})^2}}AM_T^\alpha dM_T}
	\end{equation}
for $\alpha= -1$ and $\alpha= 0$.

   	\begin{equation}
	y(\alpha= -1) = A\int_{M_0}^X{\frac{M_0/M_T^3}{\sqrt{1-(\frac{M_0}{M_T})^2}} dM_T} = A\frac{\sqrt{X^2-M_0^2}}{M_0X}
	\end{equation}

   	\begin{equation}
	y(\alpha= 0) = A'\int_{M_0}^X{\frac{M_0/M_T^2}{\sqrt{1-(\frac{M_0}{M_T})^2}} dM_T} = A'\arccos (\frac{M_0}{X}).
	\end{equation}
We can split the two zones below and above the upper limit of the brown dwarf desert, $M_{des} = 57$ \MJup, and the total area for the normalization is:

   	\begin{equation}
	\Phi = A\int_{M_0}^{M_{des}}{\frac{M_0/M_T^3}{\sqrt{1-(\frac{M_0}{M_T})^2}} dM_T} + A'\int_{M_{des}}^{80 M_J}{\frac{M_0/M_T^2}{\sqrt{1-(\frac{M_0}{M_T})^2}} dM_T},
	\end{equation}
where the constant $A'=A/M_{des}$ because we want the two curves to be connected in the value of $M_{des}$.

If the mass of detection is $M_0 \leq X \leq M_{des}$ then the probability is calculated through:

   	\begin{equation}
	P(M_T > X \mid M_0) = 1 - \frac{A\int_{M_0}^X{\frac{M_0/M_T^3}{\sqrt{1-(\frac{M_0}{M_T})^2}} dM_T}}{A\int_{M_0}^{M_{des}}{\frac{M_0/M_T^3}{\sqrt{1-(\frac{M_0}{M_T})^2}} dM_T} + \frac{A}{M_{des}}\int_{M_{des}}^{80 M_J}{\frac{M_0/M_T^2}{\sqrt{1-(\frac{M_0}{M_T})^2}} dM_T}}
	\end{equation}
that gives

   	\begin{equation}
	P(M_T > X \mid M_0) = 1 - \frac{\frac{\sqrt{X^2-M_0^2}}{M_0X}}{\frac{\sqrt{M_{des}^2-M_0^2}}{M_0M_{des}} + \frac{1}{M_{des}} \arccos (\frac{M_0}{80 M_J})}.
	\end{equation}
For $M_{des} < X \leq 80 M_J$ in the same way we obtain:

   	\begin{equation}
	P(M_T > X \mid M_0) = 1 - \frac{\frac{\sqrt{M_{des}^2-M_0^2}}{M_0 M_{des}} + \frac{1}{M_{des}} \arccos (\frac{M_0}{X})}{\frac{\sqrt{M_{des}^2-M_0^2}}{M_0 M_{des}} + \frac{1}{M_{des}} \arccos (\frac{M_0}{80 M_J})}.
	\end{equation}
In this way we obtained the probability that an object is more massive than the minimum mass value for the detection and that it is not a stellar type companion.


\section{GJ\,676 Candidate monitoring}
\label{a:cand}

\begin{center}
	\begin{figure}
	\centering
	\includegraphics[width=0.5\textwidth]{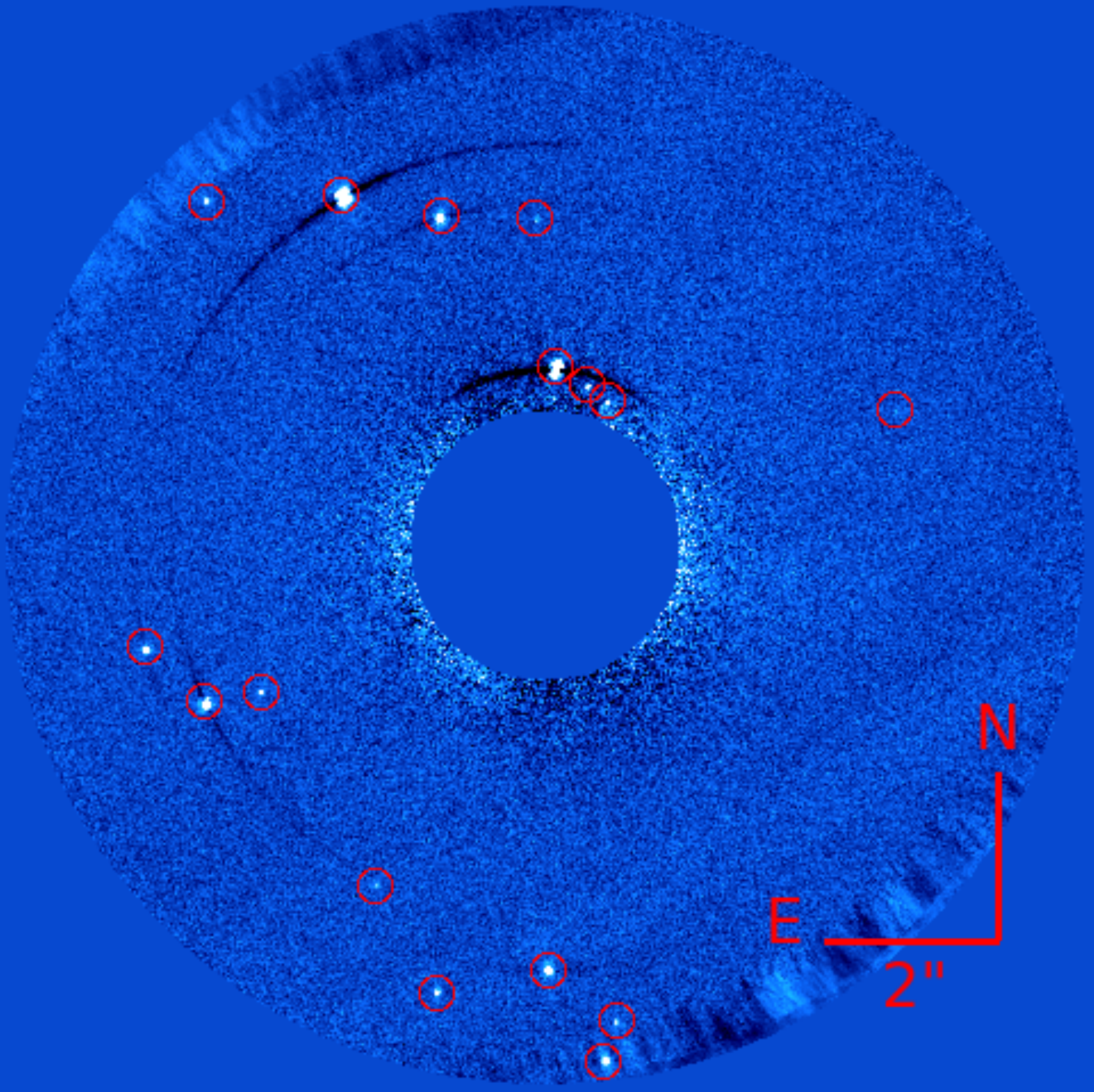}
	\caption{Background stars found in the IRDIS FOV of GJ~676. The candidates are in the red circles. }
	\label{f:gj}
        \end{figure}
\end{center}

During the first epoch of observation of the star GJ\,676, 14 companions candidates were identified with a signal to noise ratio above 5. Two other candidates have been identified with lower S/N and all 16 candidates are shown in Figure~\ref{f:gj}. The second epoch, taken after one year, confirmed that all of them are background sources.
Table~\ref{t:idstars} presents the list of the astrometric relative positions of each star with respect to GJ\,676.

\begin{table}
 \centering
\caption[]{Astrometric positions of the background stars around GJ\,676. The sky coordinates of the star for the epoch of May 2016 are RA = 17.50298568 and Dec = -51.63782484}
\label{t:idstars}
\begin{tabular}{ccc}
\hline
\hline
Star ID & $\Delta$RA (\as)   &  $\Delta$Dec (\as)  \\
\hline
1& 3.95& -2.28\\
2& 3.90& -3.84\\
3& 3.71& -1.18\\
4& 1.98&  0.12\\
5& 1.79&  0.48\\
6& 1.61&  0.70\\
7&-1.18& -4.53\\
8&-1.66& -3.22\\
9&-1.81& -3.84\\
10&-3.86& -1.91\\
11&-4.82&  0.04\\
12&-5.07& -1.22\\
13&-5.42&  0.80\\
14&-5.85&  0.69\\
\hline
\end{tabular}
\end{table}

\label{lastpage}

\end{document}